\documentclass[12pt,preprint,xdvi,brief_bib]{aastex}

\usepackage{natbib}
\usepackage{graphicx}

\newcommand{\be}{\begin{equation}}
\newcommand{\ee}{\end{equation}}
\newcommand\eq{Eq.}
\newcommand\eqs{Eqs.}
\newcommand\fig{Fig.}
\newcommand\figs{Figs.}

\newcommand{\half}{\hbox{${1\over2}$}}

\newcommand{\zhat}{{\bf \hat{z}}}

\newcommand{\vvec}{{\bf v}}
\newcommand{\bvec}{{\bf B}}
\newcommand{\evec}{{\bf E}}
\newcommand\quarter{\hbox{${1\over4}$}}

\newcommand\thh{\hbox{${3\over2}$}}

\newcommand\fhf{\hbox{${5\over2}$}}

\newcommand\kb{k_{\rm B}}
\newcommand\pr{{\rm Pr}}

\begin{document}

\title{Slow shocks and conduction fronts from Petschek reconnection of skewed magnetic fields: two-fluid effects}

\author{D.W. Longcope,$^{1,2}$ and S.J. Bradshaw$^{3,4}$}
\affil{1. University of Maryland, College Park, MD 20742\\
  2. Permanent address: Department of Physics, Montana State University, \\
  Bozeman, MT 59717\\
3. NASA Goddard Space Center Flight, Greenbelt, MD 20771\\
4. Department of Physics and Astronomy, Rice University, Houston, TX 77005}


\begin{abstract}
In models of fast magnetic reconnection, flux transfer occurs within a small portion of a current sheet triggering stored magnetic energy to be thermalized by shocks.  When the initial current sheet separates magnetic fields which are not perfectly anti-parallel, i.e.\ they are {\em skewed}, magnetic energy is first converted to bulk kinetic energy and then thermalized in slow magnetosonic shocks.  We show that the latter resemble parallel shocks or hydrodynamic shocks for all skew angles except those very near the anti-parallel limit. As for parallel shocks, the structures of reconnection-driven slow shocks are best studied using two-fluid equations in which ions and electrons have independent temperature.  Time-dependent solutions of these equations can be used to predict and understand the shocks from reconnection of skewed magnetic fields.  The results differ from those found using a single-fluid model such as magnetohydrodynamics.  In the two-fluid model electrons are heated indirectly and thus carry a heat flux always well below the free-streaming limit.  The viscous stress of the ions is, however, typically near the fluid-treatable limit.  We find that for a wide range of skew angles and small plasma $\beta$ an electron conduction front extends ahead of the slow shock but remains within the outflow jet.  In such cases conduction will play a more limited role in driving chromospheric evaporation than has been predicted based on single-fluid, anti-parallel models.
\end{abstract}

\date{Draft: \today}

\section{Introduction}

Magnetic reconnection was proposed to explain the conversion of magnetic energy to other forms in solar flares \citep{Giovanelli1947}.  In an early detailed model of reconnection 
\citep{Parker1957,Sweet1958} plasma was accelerated and heated all within the same current sheet that initially stored the magnetic energy.  Reconnection in this mode was, however, deemed too slow to explain flares.  In an alternative proposed by 
\citet{Petschek1964} the reconnection process decomposed the initial current sheet into a series of waves and shocks.  Flux transfer occurred within a very small portion of the original sheet, the diffusion region, while acceleration and heating occurs in the much larger slow magnetosonic shocks.  

While it was initially analyzed in steady-state, subsequent work has shown the Petschek scenario applies more generally.  Even a transient transfer of magnetic flux across a pre-existing current sheet will, if it occurs within a localized portion, produce the same set of shocks proposed by Petschek 
\citep{Semenov1983,Biernat1987,Heyn1996,Erkaev2000,Nitta2001}.  When the flux transfer is impulsive, the current sheet decomposes progressively as the shocks propagate away from the flux transfer point. 

In solar flare models, the heat from a slow shock is expected to be be conducted along field lines {\em ahead of} the shock itself \citep{Forbes1989,Tsuneta1996}.  Except for possible non-thermal particles \citep{Karpen1987}, this conduction front (CF) was predicted to be the first effect from reconnection energy release to reach the lower atmosphere and drive chromospheric evaporation.  A conduction front was identified by \citet{Yokoyama1997} in numerical simulations and later demonstrated to be an effective driver of evaporation 
\citep{Yokoyama1998}.  The structure of the CF in this particular simulation has been recently analyzed by \citet{Seaton2009} using a boundary layer method.

Conduction fronts preceding shocks are well-known features in hydrodynamics 
\citep{Becker1922,Thomas1944,Grad1951} and magnetohydrodynamics 
\citep[MHD,][]{Kennel1988,Xu1992}, when thermal conductivity greatly exceeds viscosity.  In an unmagnetized plasma viscous stress arises from ion momentum while heat flux is carried primarily by the faster-moving electrons.  The same is true of transport parallel to the field in a magnetized plasma.  In either case, the ratio of viscosity to thermal conductivity, the Prandtl number, is proportional to the ratio of the thermal velocities of the two species, which is the square-root of their mass ratio and hence very small.

The reasoning above assumes, as does any single-fluid hydrodynamics such as MHD, that the temperatures of both species are equal everywhere and always.  More sophisticated two-fluid  treatments, where the electrons and ions both have nearly Maxwellian velocity distributions but with separate temperatures, show the equal-temperature assumption to be invalid throughout a typical parallel shock \citep{Jukes1957,Tidman1958,Jaffrin1964}.   Solutions do include features similar to those of single-fluid models, although modified by the disparate temperatures.  There is an inner sub-shock, also called an ion shock, where ion kinetic energy is thermalized and a preceding CF where heat is carried by electrons.  In two-fluid models, however, heat must be transferred from ions to electrons at a collision frequency far smaller than those responsible for either viscosity or conduction.  This transfer occurs downstream of the two-fluid shock, in a region called the 
{\em ion cooling region}.  Owing to the lower frequency of exchange this region is far larger than the other shock regions; it is necessarily absent from single-fluid treatments.

Given the critical role played in flares by shocks and their associated CFs, it is essential to model them using a two-fluid formalism.  The two-fluid effects will depend on the magnetic field at the slow magnetosonic shock outside the flux transfer region.  This will depend in turn on the field initially separated by the current sheet --- the field lines joined together by the reconnection.  Petschek's original model applied to a current sheet separating perfectly anti-parallel fields and had two ``switch-off'' shocks across which the field direction and strength changed dramatically.  The model was subsequently extended to {\em skewed} cases where the initial current sheet separated fields differing by 
$\Delta\theta\ne180^{\circ}$ \citep{Petschek1967,Soward1982b,Skender2003}.  In this case each slow magnetosonic shock is replaced by two different shocks: a rotational discontinuity, where the field direction changes at constant magnitude and temperature, and a slow shock where kinetic and magnetic energy are partially converted to heat.  The CF will originate from the slow shock so this is where we expect ion and electron temperatures to differ in a two-fluid solution.  

Most studies of reconnection-generated CFs have used single-fluid equations (MHD) and worked with the special case of anti-parallel reconnection ($\Delta\theta=180^{\circ}$).  The present study examines the structure of CFs for the more general case, $\Delta\theta\ne180^{\circ}$, using two-fluid equations.  It turns out that for most such cases, say $\Delta\theta\la120^{\circ}$,  the slow magnetosonic shock is very similar to a parallel shock or an unmagnetized hydrodynamic shock.  We therefore perform our analysis using two-fluid equations for an unmagnetized plasma and find steady-state solutions very similar to those of \citet{Jukes1957}, \citet{Tidman1958} or \citet{Jaffrin1964}.  To apply these results to Petschek reconnection, however, it is necessary to use time-dependent solutions.  We find that such evolution is dominated by the very slow thermal equilibration between electrons and ions.  In both respects the shocks differ from single-fluid solutions.

\section{Petschek reconnection and hydrodynamic shocks}

\begin{figure}[htp]
\plotone{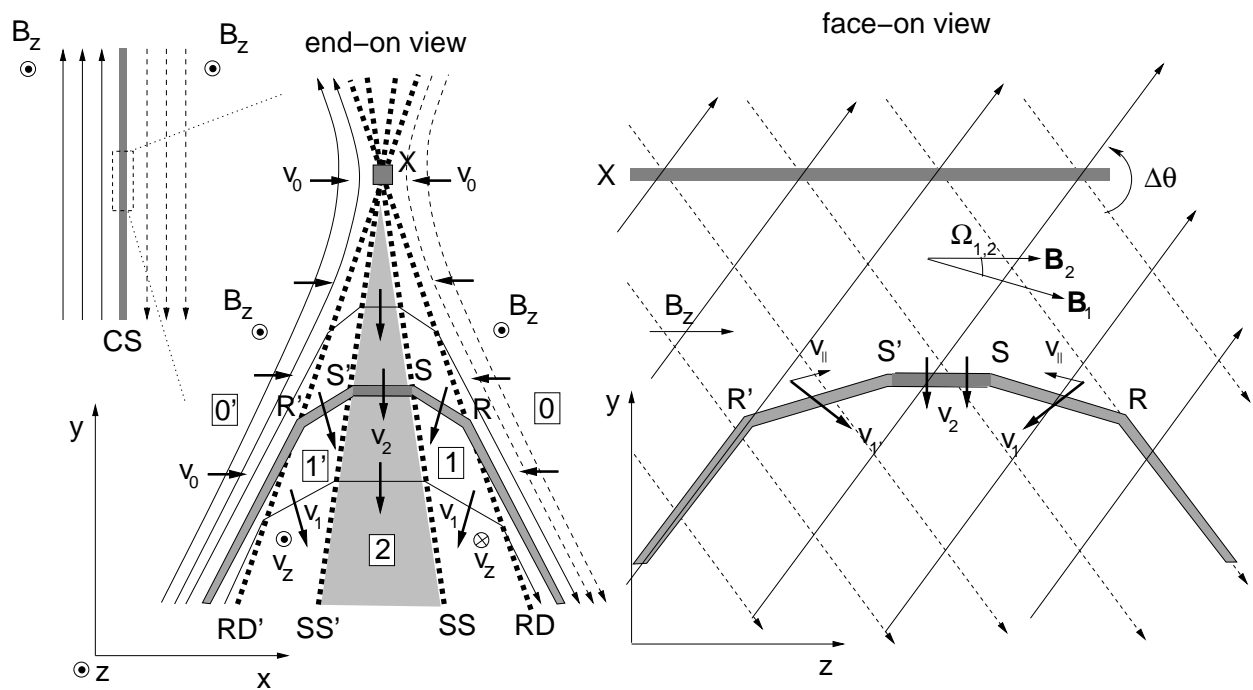}
\caption{Petschek reconnection of skewed magnetic fields.  Reconnection occurs within a portion of a large-scale current sheet (far left).  Two perspectives of the reconnection region are shown: the end-on view (center, whose horizontal scale is exaggerated for clarity) and the face-on view (right).  Flux transfer occurs along a line {\sf X} (grey) along the ignorable direction ($\zhat$).  Shocks, {\sf RD'}, {\sf SS'}, {\sf SS} and {\sf RD} (thick dashed lines) originate there, dividing the reconnection region into sub-regions labeled {\sf 0'}, {\sf 1'}, {\sf 2}, {\sf 1} and {\sf 0} (inside boxes).  The central region, {\sf 2} (shaded grey) contains plasma which has been heated and compressed by the slow shocks.  Reconnected field lines (thin solid lines) bend at each shock.  A single flux tube is called out in grey in each panel.  It passes though the shocks at points  {\sf R'}, {\sf S'}, {\sf S} and {\sf R}, left to right.}
	\label{fig:toon}
\end{figure}

Reconnection occurs across a current sheet (tangential discontinuity) separating field lines of equal magnitude, $B_0$, but differing in direction by $\Delta\theta$. The basic premise of the Petschek model is that a localized electric field occurs within an otherwise steady current sheet.  Original versions of the model attempted to attribute this electric field to Ohmic resistivity, but it has subsequently been found that any mechanism producing $\evec$ with a component along the magnetic field will generate very similar external structure --- structure outside the ``diffusion region'' designated {\sf X} in \fig\ \ref{fig:toon}.  That external structure consists of several shocks emanating from the diffusion region.  In a strictly steady-state interpretation these are standing shocks, however, they are more realistically understood as the ends of propagating shocks encompassing the field lines which were topologically changed by the inner electric field \citep{Semenov1983}.  The diffusion region can affect very little of the plasma directly, owing to its very small size, so it is the shocks launched by the reconnection which convert the magnetic energy stored in the current sheet to other forms.

Two types of MHD shocks are launched from the diffusion region, rotational discontinuities (RDs) and slow magnetosonic shocks (SSs) also called slow shocks 
\citep{Petschek1967,Soward1982b}.  
Magnetic tension accelerates the plasma at the RDs creating an Alfv\'en-speed outflow jet.  Slow shocks occur within this jet, heating and compressing the plasma.  Thus only the inner portion of the outflow jet is hot.  In the special case of anti-parallel fields ($\Delta\theta=180^{\circ}$) these two distinct shocks merge into a single slow shock of the switch-off type, often associated with the Petschek model.  Naturally one cannot assign, in this singular limit, different roles to the different shocks, so one often hears that slow shocks accelerates the outflow jet as well as heat the plasma.  In the general case, however, slow shocks only heat while RDs accelerate the jet.

There is a relatively small inward velocity, the ``inflow'' which we call 
$\vvec_0$, in the outer region (regions {\sf 0} and {\sf 0'} of \fig\ \ref{fig:toon}).  In genuine steady state models this is often interpreted as an external driver of reconnection, however, in transient or spontaneous models it is part of a fast magnetosonic rarefaction wave launched at the same time as the shocks \citep{Lin1993,Nitta2001}. In that case the flow is {\em caused by} the reconnection and does not {\em cause} the reconnection.  While it is a relatively slow flow ($v_0\ll v_{A,0}$) it has been the focus of intensive study since reconnection models were first proposed --- indeed, how it compares to $v_{A,0}$ is generally regarded as the crux of the fast reconnection problem.  In spite of this focus, many aspects of the post reconnection dynamics, including heating and energy release, are largely independent of $v_{0}$.  We focus exclusively on these energetics, occurring in regions {\sf 1} and {\sf 2}, and hereafter ignore any effects strictly dependent on the inflow $\vvec_0$.

\begin{figure}[htb]
\epsscale{0.8}
\plotone{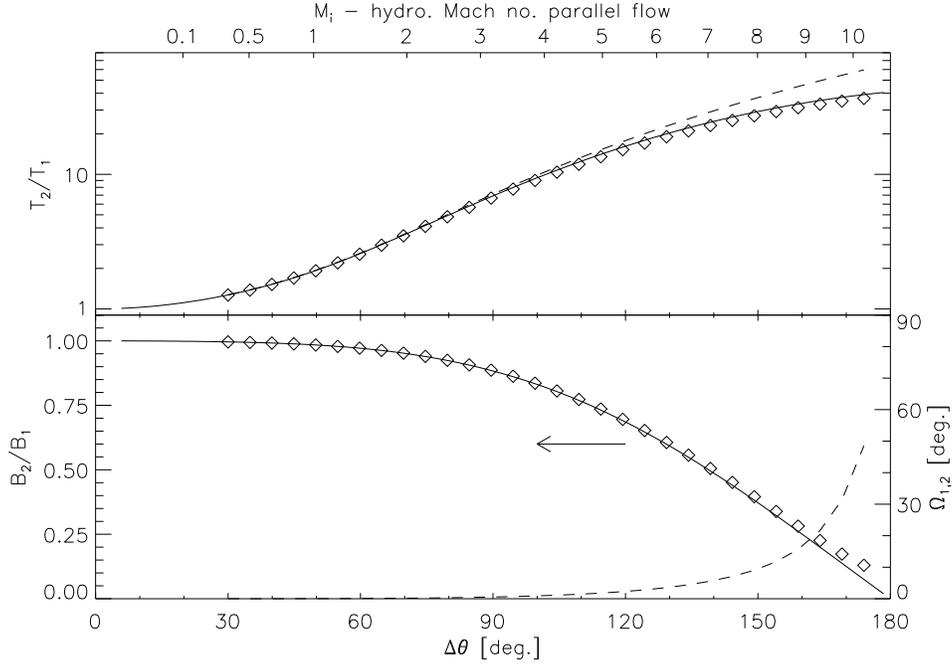}
\caption{Jumps across the slow shock as a function of reconnection angle $\Delta\theta$ for $\beta_0=0.01$.  The bottom panel shows the jump of the magnetic field magnitude, $B_2/B_1$ on the left axis, and direction, $\Omega_{1,2}$ on the right axis.  The solid curve is from the 2.5d steady model of \citet{Soward1982b}, as reported by \citet{Forbes1989} and \citet{Vrsnak2005}, and the diamonds are from a 1d Riemann problem solution of \citet{Lin1993}.  The dashed curve is the angle, read off the right axis.  The top panel shows the temperature ratio for the same two models (solid and diamonds).  The parallel flow speed, $v_{\parallel,1}$ for the 1d Riemann solution is used to compute a hydrodynamic mach number, $M_i$, plotted along the top axis.  The corresponding temperature ratio from a 1d hydrodynamics shock, is plotted as a dashed curve.}
	\label{fig:brat}
\end{figure}

The structures of the different regions follow from the standard jump conditions of the MHD shock types.  The RD is a non-linear manifestation of a shear Alfv\'en wave and thermodynamic quantities, density, pressure and temperature are all continuous across 
it \citep{Priest2000}.  Thus the RDs convert magnetic energy to bulk kinetic energy.  The magnetic field changes direction but not magnitude there so the magnetic energy release is due to field line shortening alone.   The slow shocks are the site of all heating; it is there that the temperature and density both jump to higher post-shock values, $T_2$ and $\rho_2$.   The magnitude of the field weakens across slow shock so $B_2\le B_1$.  Figure \ref{fig:brat} shows that for typical coronal conditions, $\beta_0=0.01$, the amount of weakening is modest except for the anti-parallel case where the shock can be considered a ``switch-off'' shock.  The RD rotates the magnetic field to enhance the ignorable component: $B_{z1}>B_{z0}$.  The slow shock then weakens this enhanced component, but generally leaves it greater than it started: $B_{z1}>B_{z2}>B_{z0}$.

With only a modest weakening of the magnetic field, the energy for plasma heating comes largely from bulk flow within region {\sf 1}.  The RDs create a flow component $v_{\parallel,1}$ along the local magnetic field.  To see this consider the RD from a frame moving 
with point {\sf R} in \fig\ \ref{fig:toon}.  This point moves along the unperturbed magnetic field at speed $v_{\rm A,0}$.   In this co-moving frame the plasma flows parallel to the field line at constant speed changing direction by the angle $\Omega_{0,1}$ ---  the angle between 
$\bvec_0$ and $\bvec_1$.  Neglecting the small external inflow ($\vvec_0\simeq0$) the flow speed in the co-moving frame is that of the reference frame,  $v_{\rm A,0}$, on both sides of the RD.  Returning to the lab frame we find the parallel component in region {\sf 1} to be
\be
  v_{\parallel,1} ~=~ v_{\rm A,0} - v_{\rm A,0}\cos\Omega_{0,1} ~=~
  2v_{\rm A,0}\sin^2(\Omega_{0,1}/2) ~~.
  	\label{eq:vpar}
\ee
While the Lorentz force cannot accelerate parallel to the local field, the net effect of a bend is acceleration along its bisector.  This produces flow with a component parallel to the field on the non-stationary side.  Indeed, such inward parallel flow is necessary for shortening the magnetic field lines.

When the external plasma is characterized by very small $\beta_0$, the parallel flow speed is typically greater than the sound speed,
\be
  M_{i} ~=~ {v_{\parallel,1}\over c_{s,1}} ~=~ 
  2{v_{\rm A,0}\over c_{s,0}}\sin^2(\Omega_{0,1}/2) ~=~
  \sqrt{{8\over\gamma\beta_0}} \sin^2(\Omega_{0,1}/2) ~~,
  	\label{eq:Mi}
\ee
where $\gamma=5/3$ is the ratio of specific heats.  When $\beta_0\ll1$ each RD produces a flow at or above Mach 1, directed toward the center.  More than any magnetic change, it is the collision of these supersonic flows that accounts for the slow shocks.  Indeed, \fig\ \ref{fig:brat} shows the magnetic field changes direction very little across these shocks so they are akin to parallel shocks.  Except for nearly anti-parallel cases 
$\Delta\theta\simeq180^{\circ}$, the SSs are essentially hydrodynamic shocks from pistons 
moving $M_{i}$ times the ambient sound speed.  The top axis in \fig\ \ref{fig:brat} shows $M_i$ corresponding to the current sheet angle, $\Delta\theta$, and the dashed curve is the temperature ratio for a one-dimensional hydrodynamic shock at that Mach number 
\citep{Courant1948}.

\begin{figure}[htb]
\epsscale{0.8}
\plotone{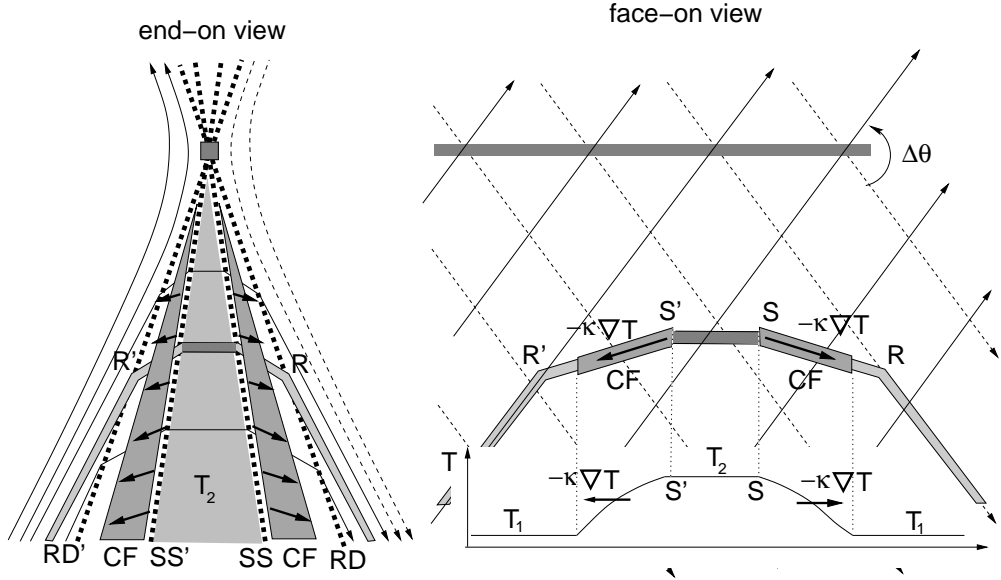}
\caption{Expected location of the conduction fronts ({\sf CF}) ahead of slow shocks.  Reconnection configuration is the same as \fig\ \ref{fig:toon}.  Additional grey  regions represent the CFs and arrows within them show the direction of heat flux.  An inset along the bottom of the {\sf face-on view} shows the distribution of temperature along the flux tube.  It has constant value, $T_2$ within region {\sf 2} and decrases smoothly to $T_1$ through the CF.  The temperature gradient drives the heat flux, $-\kappa\nabla T$, as indicated by the arrows.}
	\label{fig:cf_toon}
\end{figure}

The jump conditions across MHD shocks are dictated by conservation of mass, momentum and energy through the Rankine-Hugoniot conditions.  Transport effects, such as viscosity and conductivity, conserve all of these and therefore do not modify jump conditions.  They do, however, determine the ``inner structure'' of the jumps including the length scale over which the quantities change 
\citep{Kennel1988}.  The largest of the transport coefficients is thermal conduction parallel to the magnetic field.  This conducts heat from region {\sf 2} into pre-shock regions {\sf 1'} and 
{\sf 1}, creating a {\em conduction front} (CF) ahead of the slow shocks as shown in \fig\ \ref{fig:cf_toon}.  It is not immediately clear under what conditions these fronts might extend far enough to reach beyond the RDs.  They clearly must for anti-parallel fields since the RD and SS coincide --- the only circumstance in which Petschek-driven CFs were previously studied.

Instead of tackling this problem using the full equations of MHD we make use of the close relation, for most skewed fields, between the slow shocks and hydrodynamic shocks.  Since transport effects do not modify the jump conditions we know that cases where 
$\Delta\theta\la120^{\circ}$ will involve collision along a flux tube bending by not more than $3^{\circ}$ across the SS.  This is so close to being a straight tube that we apply the analysis of a piston in a straight shock tube.  

The initial energy in such a system is primarily the bulk kinetic energy of the  ions composing the super-sonic flow.  Its deceleration at the shock converts most of this kinetic energy to thermal energy --- thermal energy of ions.  The ions must then transfer some of their heat to the electrons whose much higher conductive flux creates the CFs.  It is thus apparent that the internal structure of a hydrodynamic shock must be analyzed in terms of separate ion and electron fluids.  Steady-state analyses of this kind were performed by 
\citet{Jukes1957}, \citet{Tidman1958} and \citet{Jaffrin1964}.  We repeat this in \S 4, below, adapting it slightly, before analyzing the transient behavior in \S 5.  Owing to the slowness with which ion heat can be transferred to electrons, the steady state would be achieved only gradually if ever.  In \S 6 we place the one-dimensional, time dependent shock solutions into a
steady-state Petschek reconnection jet to ascertain the effects of two-fluid equations on slow shocks and CFs.

\section{The Shock tube problem}

To model the structure and development of the slow shock we solve for hydrodynamic flow into a straight shock tube with a fixed end at $s=0$, representing the center of region {\sf 2}.  The fluid enters the tube from the right ($s>0$) with properties of the ambient, pre-reconnection plasma, $\rho_1$ and $T_1$, (none of them were changed at the RD). Although we treat the electron and ion fluids separately, we assume they are in equilibrium within the inflow, 
$T_{e1}=T_{i1}=T_1$.  The flow enters the tube with a leftward velocity 
$v=-v_{\parallel,1}$ fixed by the RD according to \eq\ (\ref{eq:vpar}).  The flow actually originates from the RD, point {\sf R}, receding at speed $v_{\rm A,0}$.  We assume that this outruns any effect from region {\sf 2} and take the uniform inflow region to extend to $s\to+\infty$.  

In the view of this shock tube picture we use the term {\em inflow} for the inward flow, 
$v=-v_{\parallel,1}$, originating from the open end of the shock tube.  It should be emphasized that it represents a component of $\vvec_1$ in \fig\ \ref{fig:toon}, which is itself inside what is often termed the {\em outflow jet}.  More significantly, it is not the same as the {\em reconnection inflow}, $\vvec_0$, which goes {\em into the outflow jet from outside}.  Nor is our shock-tube Mach number, $M_i$, related to the Alfv\'en Mach number often used in reconnection theory.  Hereafter we consider only one-dimensional hydrodynamics in a straight tube.

\subsection{Two-fluid equations}

For the reasons elaborated above we use separate equations for ion and electron fluids. We assume the shock thickness will be much larger than the  Debeye length so quasi-neutrality is appropriate: $n_e=n_i=n$.  The central region is current-free so we assume no current crosses the slow shock which means that the charged fluids move together: $v_e=v_i=v$ along the tube (i.e.\ normal to the shock).  The common density and velocity satisfy continuity and the ion momentum equation
\begin{eqnarray}
  {\partial n\over \partial t} &=& - {\partial\over\partial s}(nv) \label{eq:cont} \\
    {\partial (m_i nv)\over \partial t} &=& - {\partial\over\partial s}(m_inv^2)
    -  {\partial\over\partial s}(p_i + p_e) ~+~ {\partial\over\partial s}\left({4\over 3}
    \mu_i{\partial v\over \partial s}\,\right) ~~, \label{eq:mom_raw}
\end{eqnarray}
where $m_i$ is the mass of an average ion (assumed to be singly charged) and
$\mu_i$ is the dynamic viscosity of the ion fluid.  The electron pressure appears in place of the electric field after neglecting inertia in the electron momentum equation
\be
  eE_s ~=~ -{1\over n}\,{\partial p_e\over \partial s}  ~~.
  	\label{eq:Es}
\ee
Since the electron momentum equation lacks an electron viscosity contribution that effect is absent from the ion momentum equation, \eq\ (\ref{eq:mom_raw}).  Were it included it would have been smaller than the ion viscosity by $m_e/m_i\ll1$.  It has been found by \citet{Jaffrin1964} to have an effect only within a Debeye-length layer of charge separation at the leading edge of the CF.

We assume sufficient collisionality within each species to isotropize their distributions leading to scalar pressures, $p_e=n\kb T_e$ and $p_i=n\kb T_i$ ($\kb$ is Boltzmann's constant).  The collisional energy exchange between ions an electrons,  at frequency $\nu_{ei}$, is slow enough that we consider their temperatures to be independent and use two separate energy equations.  The electron fluid's contribution to bulk kinetic energy is neglected, so its entire volumetric energy density is the internal energy
$\thh p_e$.  This evolves according to
\be
  {\partial\over \partial t} (\thh p_e) ~=~ - {\partial\over\partial s}(\fhf vp_e)
  + v{\partial p_e\over \partial s}
   +{\partial\over\partial s}\left(\kappa_e{\partial T_e\over \partial s}\,\right) 
   + \thh\kb\, n \nu_{ei}(T_i-T_e)~~, \label{eq:Te0}
\ee
where $\kappa_e$ is the electron thermal conductivity.  The second term on the right hand side (rhs) represents work done by the electric field, which has been replaced using (\ref{eq:Es}).  The third term is heat transport by thermal conduction and the final term is heat supplied by the ions. 

The scalar ion pressure evolves according to a similar equation.  To facilitate the derivation of conservation laws we write the equation governing the evolution of the total ion energy, thermal plus kinetic,
\begin{eqnarray}
  {\partial \over \partial t}\left( \thh p_i + \half m_i n v^2\right) 
  &=& - {\partial\over\partial s}\left( \fhf vp_i + \half m_i n v^3 \right)
  - v{\partial p_e\over \partial s} +  {\partial\over\partial s}\left({4\over 3}v\,
    \mu_i{\partial v\over \partial s}\,\right) \nonumber \\
    &~& + {\partial\over\partial s}\left(\kappa_i{\partial T_i\over \partial s}\,\right) 
   + \thh\kb\, n \nu_{ei}(T_e-T_i)~~. \label{eq:e_erg}
\end{eqnarray}
The second term on the rhs is work from the electric field, with sign opposite the term in \eq\ (\ref{eq:e_erg}) since the ions are positively charged.  The third term is the work done by viscous stress both on the bulk kinetic energy and on the internal energy as irreversible heating.  The final term is the heating from electrons; this is naturally equal and opposite to the corresponding term in \eq\ (\ref{eq:Te0}) since the energy is merely exchanged between the two species.

To bring out the fundamental scalings in the shock solutions we rescale all quantities using properties of the inflow.  The velocity is scaled to sound speed $c_{s1}$ which we define using the single-fluid form
\be
  c^2_{s1} ~=~ {5\over 3}\, {p_{e1}+p_{i1}\over n_1m_i} ~=~ {10\over3}
  {\kb T_1\over m_i} ~~.
\ee
The asymptotic inflow velocity, $v\to -M_i$, as $s\to+\infty$, is the one free parameter in the problem.  We rescale temperatures to $m_ic_{s1}^2/\kb$ so the normalized inflow temperature is $T_1=0.3$.  We rescale number density to $n_1$.    

The fundamental length scale in a shock is set by the viscosity.  This varies with temperature so we use the ambient value to define a reference length scale
\be
  \ell_{i1} ~=~ {4\over 3}\,{\mu_{i1}\over m_i\, n_1c_{s1}} ~\simeq~ 5.85\,{\rm km}
  \left({T_1\over 10^6\,{\rm K}}\right)^2\,\left({10^{10}\,{\rm cm}^{-3}\over n_{e1}}\right)~~,
\ee
related to the pre-shock ion mean free path.  Times are rescaled to 
$\tau_{i1}=\ell_{i1}/c_{s1}$ which is proportional to the ion-ion collision time.  

The local viscosity is proportional to the five-halves power of the local ion temperature, which we denote by $\chi_i$ for notational convenience
\be
  \mu_i ~=~ \mu_{i1}\left({T_i\over 0.3}\right)^{5/2} ~=~\mu_{i1}\chi_i(T_i) ~~.
\ee
Thermal conductivities have identical temperature dependence which we express using the same notation
\be
  \kappa_{e} ~=~ \kappa_{e1}\left({T_e\over 0.3}\right)^{5/2} =~\kappa_{e1}\chi_{e}(T_{e}) ~~,~~
    \kappa_{i} ~=~ \kappa_{i1}\chi_{i}(T_{i})~~.
\ee
The ratios of viscosity to the different conductivities defines Prandtl numbers
\be
  \pr_{e,i} ~=~ {4\over 3}\,{\mu_{i}\over \kb\kappa_{e,i}} ~~;
\ee
which are independent of temperature.  For Spitzer values of these transport coefficients, and taking $m_i$ to be the proton mass, these dimensionless quantities are $\pr_e=0.013$ and $\pr_i=0.33$.
The electron-ion collision frequency has a related temperature dependence, which we write as
\be
  {\nu_{ei}\over n} ~=~ {4\sqrt{2\pi}\,e^4\over 3m_em_i}
  \left({\kb T_e\over m_e}\right)^{-3/2}\, \ln\Lambda ~=~ 
  {2\over 3}\,{f_1\,\chi_e^{-3/5}\over n_1\tau_{1,i}} ~~,
\ee
where $\ln\Lambda$ is the Coulomb logarithm.  For Spitzer collision rates the coefficient is $f_1=0.019$.

The rescaling yields a set of equations including an unchanged continuity 
equation, (\ref{eq:cont}), and
 \begin{eqnarray}
    {\partial (nv)\over \partial t} &=& - {\partial\over\partial s}(nv^2)
    -  {\partial\over\partial s}[n(T_i + T_e)] ~+~ {\partial\over\partial s}
    \left[\chi_i{\partial v\over \partial s}\,\right] \label{eq:mom} \\
  {\partial (\thh nT_e)\over \partial t} &=& - {\partial\over\partial s}(\fhf vnT_e)
  + v{\partial (nT_e)\over \partial s} +{\partial\over\partial s}\left[{\chi_e\over \pr_e}
  \left({\partial T_e\over \partial s}\right) \right]     \nonumber \\
  &~& + f_1 n^2\chi_{e}^{-3/5}(T_i-T_e)~~, \label{eq:Ee} \\
  {\partial \over \partial t}\left( \thh nT_i + \half n v^2\right) 
  &=& - {\partial\over\partial s}\left( \fhf vnT_i + \half n v^3 \right)
  - v{\partial (nT_e)\over \partial s} +  {\partial\over\partial s}
    \left(v\,\chi_i{\partial v\over \partial s}\,\right) \nonumber \\
    &~& + {\partial\over\partial s}\left({\chi_i\over\pr_i}{\partial T_i\over \partial s}\,\right) 
   - f_1 n^2 \chi_e^{-3/5}(T_i-T_e)~~. \label{eq:Ei}
\end{eqnarray}
These are to be solved with inflow boundary conditions, $v=-M_i$, $T_e=T_i=0.3$ and $n=1$ at the right, $s\to+\infty$ and mirror conditions at $s=0$: $v=0$, 
$\partial T_e/\partial s=\partial T_i/\partial s=\partial n/\partial s=0$.  The only free parameter, $M_i$, in the inflow boundary condition, represents through \eq\ (\ref{eq:Mi}), the Alfv\'en speed and current sheet angle $\Delta\theta$ of the reconnection.  

\subsection{The single-fluid limit}

It is the small value of the dimensionless constant $f_1$ that dictates the slow coupling between the electron and ion temperatures.  In spite of this fact, it is common to treat shocks using single-fluid equations whereby the two temperatures are equated, $T_e=T_i=T$.  In this case the momentum equation, (\ref{eq:mom}) becomes
\be
    {\partial (nv)\over \partial t} ~=~ - {\partial\over\partial s}(nv^2)
    -  2{\partial\over\partial s}(nT) ~+~ {\partial\over\partial s}
    \left(\chi{\partial v\over \partial s}\,\right)
\ee
where $\chi(T) = (T/0.3)^{5/2}$.  The net energy equation comes from summing (\ref{eq:Ee}) and (\ref{eq:Ei}),
\be
   {\partial \over \partial t}\left( 3nT + \half n v^2\right) 
  = - {\partial\over\partial s}\left( 5vnT + \half n v^3 \right)
   + {\partial\over\partial s}\left(v\,\chi{\partial v\over \partial s}\,\right) 
   + {\partial\over\partial s}\left({\chi\over\pr}{\partial T\over \partial s}\,\right)
\ee
where the combined Prandlt number, $\pr=\Pr_e\Pr_i/(\Pr_i+\Pr_e)=0.012$, includes both electron and ion conductivities.  The pressure, $p=2nT$, is a combination of ion and electron contributions now required to be equal.  We use these single-fluid equations for comparison to a larger body of work on purely hydrodynamic shocks and MHD parallel shocks.

\section{Stationary shock solutions}

We begin the analysis of \eqs\ (\ref{eq:cont}) and (\ref{eq:mom})--(\ref{eq:Ei}) by seeking solutions which are stationary in a reference frame moving to the right at some speed $v_s$.  Since all the equations are invariant under Galilean transformation, the only modifications are to the boundary conditions.  The upstream conditions, $s\to+\infty$, are the same, $T_e=T_i=0.3$ and $n=1$, but with a boosted velocity 
\be
  v \to v_1 = -(v_s+M_i) ~=~ -M_1 ~~.
\ee
At the left boundary, now at $s\to-\infty$, we assume all quantities approach uniform values, $T_e\to T_{e2}$, $T_i\to T_{i2}$, $n\to n_2$ and $v\to v_2$.  (We do not assume 
{\em a priori} that the post-shock electron and ion temperatures are equal but show below that they are.)  These post-shock values must be found from the solution of the equations.  Once found, the velocity of the reference frame is then determined by making the downstream velocity 
vanish in the laboratory frame: $v_s=|v_2|$ and $M_1 = M_i+|v_2|$.

The solution we seek is time-independent in this reference frame, so $\partial/\partial t=0$ in all equations.  The continuity equation, \eq\ (\ref{eq:cont}), is then quickly integrated to yield $nv=n_1v_1=-M_1$.  This can be used to formally eliminate $n$ in favor of $v$.  The momentum equation, \eq\ (\ref{eq:mom}), can also be integrated to yield
\be
  v + {T_e+T_i\over v} +{\chi_1\over M_1}{\partial v\over \partial s} ~=~
   - M_1 - {2T_1\over M_1} ~~, 
   	\label{eq:vs1}
\ee
after dividing by $n_1v_1$ and noting that $\partial v/\partial s\to0$ in the upstream region.  A third conservation law, conservation of total energy, results from the addition of \eqs\ (\ref{eq:Ee}) and (\ref{eq:Ei})
\be
  \fhf (T_e+T_i) + \half v^2+{v\chi_i\over M_1} {\partial v\over \partial s} 
  + {\chi_i\over M_1\pr_i}{\partial T_i\over \partial s}
  +{\chi_e\over M_1\pr_e}{\partial T_e\over \partial s}
  = 5T_1 + \half M_1^2 
  	\label{eq:Tis1}
\ee

Equations (\ref{eq:vs1}) and (\ref{eq:Tis1}) can be cast as first order equations for the profiles $v(s)$ and $T_i(s)$.  The first is
\be
  {dv\over ds} ~=~ -{M_1\over \chi_i v}\left[ T_e + T_i + v^2 + 
  \left(M_1 + {0.6\over M_1}\right)v \right] ~=~
  -{M_1\over \chi_i v}\left[ \,T_e + T_i - 2g(v)\,\right] ~~,\label{eq:vs}
\ee
after using $T_1=0.3$.  To cast \eq\ (\ref{eq:Tis1}) as an explicit  first order equation for $T_i$  we eliminate $dv/ds$ using \eq\ (\ref{eq:vs}) and introduce the heat flux variable
\be
  F_e ~=~ {\chi_e\over\pr_e}{d T_e\over ds} ~~, \label{eq:Tes}
\ee
yielding
\begin{eqnarray}
  {dT_i\over ds} &=& -{M_1\pr_i\over\chi_i}\left[ \thh(T_e+T_i) +
  \half v^2 + 2g(v) -\half M_1^2 - \thh + F_e/M_1 \right] \nonumber\\
  &=& -{3M_1\pr_i\over 2\chi_i}\left[ T_e+T_i - 2h(v) + \hbox{${2\over3}$}F_e/M_1 \right] ~~.
  \label{eq:Tis}
\end{eqnarray}
For notational convenience we have introduced two quadratic functions of velocity
\be
  g(v) ~=~ -\half v\left(v+M_1 + 0.6/ M_1\right) ~~,~~
  h(v) ~=~\hbox{${1\over 6}$}\left[ v^2 + 2(M_1 + 0.6/M_1)v + M_1^2 + 3 \right] ~~.
  	\label{eq:gh}
\ee

A final relation must come from either the electron or ion energy alone.  Neither of these is independently conserved, so the second-order equation can not be simply integrated.  The electron equation, (\ref{eq:Ee}), is, however, of first order in heat flux $F_e$.  Using \eq\ (\ref{eq:vs}) to once again eliminate $dv/ds$ gives the explicit equation
\be
  {dF_e\over ds} = -{3M_1\pr_e\over2\chi_e}F_e ~+~ {M_1^2\over \chi_i}
  {T_e\over v^2}\,[ T_e + T_i - 2g(v) ] ~+~
  {M_1^2f_1\over \chi_e^{3/5} v^2}\,( T_e - T_i )
  ~~. \label{eq:Fes}
\ee

Equations (\ref{eq:vs})--(\ref{eq:Fes}) are four first-order equations for the four profiles $v(s)$, $T_i(s)$, $T_e(s)$ and $F_e(s)$.  Their solution describes the smooth transition from an upstream state, characterized by $v=-M_1$, $T_e=T_i=0.3$,  
and $F_e=0$, to the post-shock state down stream.  These two states must be fixed-point solutions of the equations: solutions for which $d/ds=0$.  

\subsection{Fixed points}

Fixed point solutions to \eqs\ (\ref{eq:vs})--(\ref{eq:Fes}) are found by requiring $d/ds=0$ in each equation.  According to \eq\ (\ref{eq:Tes}) any fixed point state must have vanishing electron heat flux, $F_e=0$.   Equation (\ref{eq:vs}) requires that $T_e+T_i=2g(v)$ at the fixed point.  
Using these two in \eq\ (\ref{eq:Fes}) shows that $T_e=T_i$ at a fixed point solution.  Using all of these in \eq\ (\ref{eq:Tis}) shows that $T_e=T_i=h(v)$.  Since $T_i=g(v)=h(v)$, the velocity must be a root of the quadratic equation
\be
  h(v) ~-~g(v) ~=~ 0 ~=~ \hbox{${1\over 6}$}(v+M_1)( 4v + M_1 + 3/M_1) ~~.
\ee
There are therefore exactly two fixed point solutions to the coupled set of four differential equations (\ref{eq:vs})--(\ref{eq:Fes}).  These are naturally the upstream and downstream state between which the shock makes a continuous transition.

One solution of the quadratic equation, $v=-M_1$, is the trivial one: the upstream state used to set the constants in the first place.  The other fixed point solution,
\be
  v_2 ~=~ -\quarter M_1 ~-~ \hbox{${3\over 4}$}M_1^{-1} ~~,
\ee
is the downstream state.   The downstream temperature is $T_2=g(v_2)$.  The ratios of fixed point quantities can be seen to match the well-known Rankine-Hugiot relations for single-fluid hydrodynamics
\begin{eqnarray}
  {v_2\over v_1} &=& \quarter ~+~ \hbox{${3\over 4}$}M_1^{-2}
  ~=~  {n_1\over n_2} ~~,\label{eq:vrat} \\
  {T_2\over T_1} &=& {5\over 16}\left(M_1+{3\over M_1}\right)
  \left(M_1-{1\over5M_1}\right) \label{eq:Trat}
\end{eqnarray}
Thus it seems that a two-fluid shock satisfies the same Rankine-Hugoniot conditions as a simple hydrodynamic shock.

From these we find an explicit expression for the inflow Mach number
\be
  M_i ~=~ M_1 - |v_2| ~=~ \hbox{${3\over 4}$}(M_1 - M_1^{-1}) ~~.
\ee
The dashed line in the top panel of \fig\ \ref{fig:brat} shows $T_2/T_1$, from 
\eq\ (\ref{eq:Trat}), versus this inflow Mach number, $M_i$ (the top axis).

The internal shock structure is a solution of \eqs\ (\ref{eq:vs})--(\ref{eq:Fes}) diverging away from one fixed point and then converging to the other.  Finding this solution in practice requires a knowledge of the asymptotics of each fixed point solution.  The asymptotic behavior can be obtained from a linearization of the governing equations about the fixed point.  Electron and ion temperatures match at each fixed point so the two temperature dependent-functions match: $\chi_i=\chi_e=\chi$.  Using this fact we find the linearized equations
\be
  {d\over ds}\left[\begin{array}{c} \delta v \\ \delta T_i \\ \delta T_e \\
  {\delta F_e\over M_1} \end{array} \right] ~=~ {M_1\over\chi}\left[\begin{array}{cccc}
  2g'/v & -1/v & -1/v & 0 \\
  3\pr_ih' & -1.5\pr_i & -1.5\pr_i & -\pr_i \\
  0 & 0 & 0 & \pr_e \\ 
  -\displaystyle{2Tg'\over v^2} & 
  {\displaystyle{3-10f_1\over 3v^2}}T & 
  {\displaystyle{3+10f_1\over 3v^2}} T & -1.5\pr_e
  \end{array} \right]\cdot
  \left[\begin{array}{c} \delta v \\ \delta T_i \\ \delta T_e \\
  {\delta F_e\over M_1} \end{array} \right] ~~,
  	\label{eq:lin}
\ee
where all quantities in the matrix are evaluated at the fixed point.  The asymptotics near fixed point $j$ ($j=1$, 2) depends on the four eigenvalues of the matrix which we designated 
$\xi^{(j)}_1$ --- $\xi^{(j)}_4$ is ascending order.  We find that all four eigenvalues are real for both fixed points for all Mach numbers we have examined.

\subsection{Stationary solutions}

We find that the upstream fixed point has three negative eigenvalues and a single positive eigenvalue, $\xi_4^{(1)}$.  The solution can therefore converge to the upstream fixed point along any direction which combines the three eigenvectors with negative eigenvalues; this direction must be orthogonal to the fourth direction, otherwise it would diverge from the fixed point as $s\to+\infty$.   Were we to integrate \eqs\ (\ref{eq:vs}) in (\ref{eq:Fes}) backwards in $s$, {\em beginning} at the upstream point, we would be able to begin along any of the convergent direction; these span a three-dimensional sub-space.

The situation is slightly better constrained at the downstream fixed point at which we find two positive and two negative eigenvalues for all $M_1>1$.  The $s$-integration can be most easily performed from this point in the manner of a {\em shooting method} \citep{Press1992}.  Begin at a state displaced from fixed point 2 in a direction mixing, with some angle $\phi$, the two eigenvectors with positive eigenvalues.  The non-linear system, \eqs\ (\ref{eq:vs})--(\ref{eq:Fes}), are then integrated up to a state in the neighborhood of fixed point 1.  The small displacement from that fixed point is then contracted with the (left) eigenvector of the diverging direction ${\bf e}^{(1)}_4$ to produce a scalar function $Z(\phi)$.   In order for the solution to ultimately converge to the upstream state we seek a solution orthogonal to ${\bf e}^{(1)}_4$, meaning we want $Z(\phi)=0$; this will not be true in general.  The procedure is iterated, repeatedly integrating from fixed point 2 with different mixing angles $\phi$, until one is found that satisfies $Z(\phi)\simeq0$.

Through this non-linear shooting process it is possible to find a complete solution for any given Mach number $M_1$.  Following pioneering works of \citet{Thomas1944} and 
\citet{Jukes1957}, we find it useful to integrate in velocity rather than position, writing
\[
  {dx\over dv}~=~{1\over dv/dx} ~~,~~{dT_e\over dv} ~=~{dT_e/dx\over dv/dx}
  ~~,
\]
and so forth.  The system is then integrated from $v=v_2$ to $v=-M_1$.

\begin{figure}[htp]
\epsscale{1.0}
\plotone{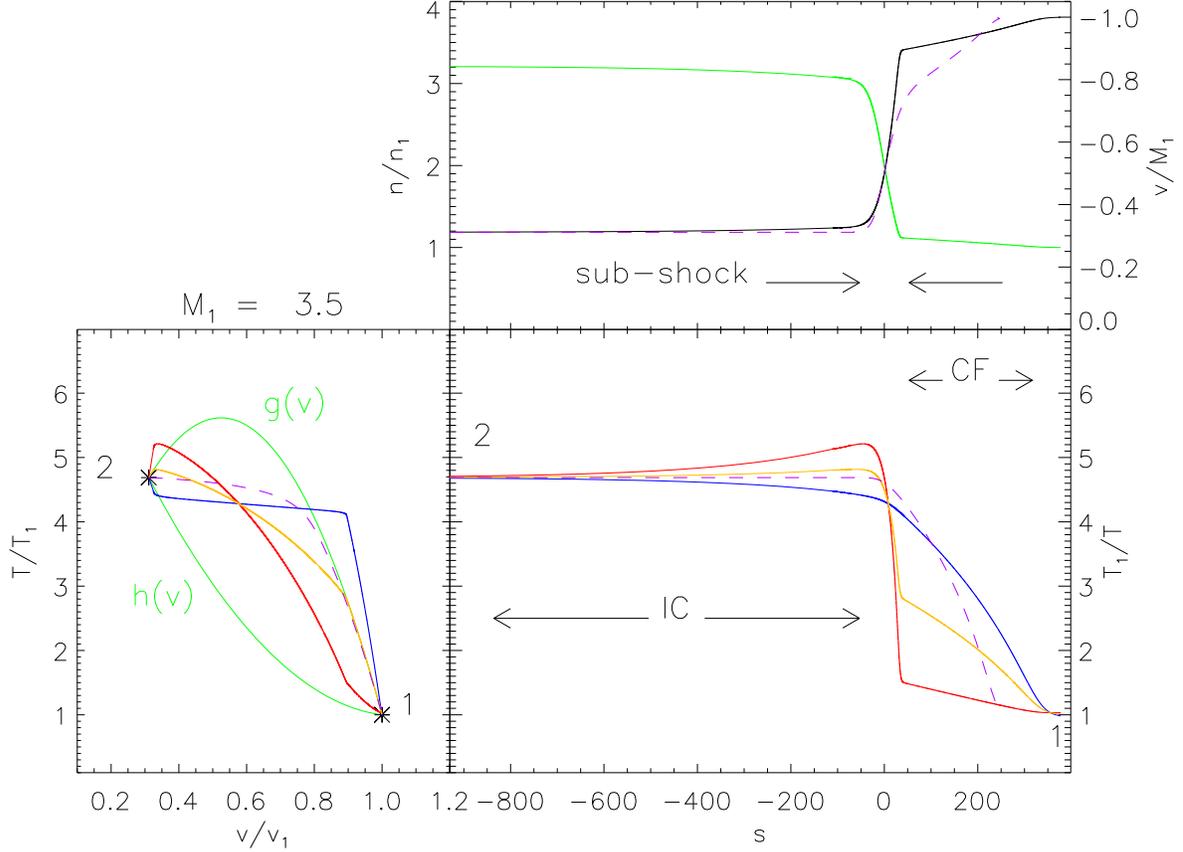}
\caption{Stationary shock profile for $M_1=3.5$.  The bottom panels show temperatures for 
electrons (blue), ions (red), their mean (orange) and the single-fluid solution (dashed violet).   The left panel shows the
phase space, $T$ {\em vs.} $v$, both normalized to upstream values.  The fixed points are
shown with asterisks and labeled {\sf 1} and {\sf 2}.  Green curves show the quadratics, $g(v)$ and $h(v)$ from \eq\ (\ref{eq:gh}).  The right panel shows the profile against distance, $s$, in which flow is from the right.  The top panel shows the velocity (black against right axis --- negative is up) and density (green against left axis).  The profiles are broken into regions which are, from right to left, the conduction front ({\sf CF}), the sub-shock and the ion cooling region ({\sf IC}).}
	\label{fig:phase}
\end{figure}

Figure \ref{fig:phase} shows the numerical solution for $M_1=3.5$ corresponding to an inflow Mach number, $M_i=2.4$; it is typical of solutions we find at other Mach numbers.  It approaches the downstream fixed point ({\sf 2}) along a direction where electron and ion  temperature discrepancies, $\delta T_e$ and $\delta T_i$, have opposite signs and are much greater than the velocity perturbation, $\delta v$.  We find this to be very close to the eigenvector with the smaller of the two positive eigenvalues, $\xi^{(2)}_3$.  This portion of the shock, where $T_i>T_e$, is called the {\em ion cooling} region.

The two temperatures cross at some point in phase space and then approach the upstream fixed point (1) with $T_e>T_i$.  It is evident from \eq\ (\ref{eq:vs}) that a monotonic velocity profile requires the mean temperature, $\bar{T}=(T_e+T_i)/2$, to be less than $g(v)$.  This means that the orange curve in \fig\ \ref{fig:phase} must remain below the upper green curve.  We find that the mean temperature, $\bar{T}$ approaches $g(v)$ from beneath with the electron temperature relatively constant.   It then follows 
$\bar{T}\simeq g(v)$ as the electron temperature decreases to $T_1$.  This portion of the solution is the {\em conduction front}, designated {\sf CF} in the figure.

The structure of the solution can be understood in terms of dynamics of the electron and ion fluids.  Approaching super-sonically from the right the ions slow abruptly to sub-sonic in the sub-shock, also called the ion shock.  Ion viscosity smooths the jump, so its size is governed by $\mu_i(T)$.  The kinetic energy of the ions is converted to ion thermal energy through viscous heating.  Since ion conduction is quite similar to the viscosity ($\pr_i=0.33$) heat flux does not broaden the sub-shock appreciably.

Quasi-neutrality demands that the electrons slow down with the ions, within the sub-shock.  Due to their much smaller mass, they carry far less kinetic energy than the ions, and the sub-shock does not heat them very much.  Since only the ions are heated at the sub-shock the electrons are cooler than the ions downstream of it.  The ions then transfer their heat to the electrons over the ion cooling region {\em via} the small collision frequency $\nu_{ei}$.  Due to their very large conductivity the electrons conduct this heat back upstream producing a CF of pre-heated electrons in advance of the sub-shock.  Its extent reflects the large electron conductivity, $\pr_e^{-1}=77$.  The electron pressure in this front partially decelerates the ions in advance of the sub-shock, through a polarization electric field.

Since the solution near fixed point {\sf 2} follows the slower of the two positive eigen-directions, $\xi^{(2)}_3$, the size and structure of the ion cooling region is dictated by it.  In the hypothetical absence of all electron-ion collisions,  $\nu_{ei}=f_1=0$, the downstream state could be a fixed point with different ion and electron temperatures provided their mean was $g(v_2)$.  This indeterminacy gives rise to a zero eigenvalue in the matrix for that hypothetical limit ($f_1=0$).  It is evident from inpsection that setting $f_1=0$ in the matrix renders it singular with null eigenvector $[0,1,-1,0]$.  Adding this to a solution would change the ion and electron temperatures without changing their mean.  The null vector is a manifestation of the temperature independence in the absence of collisions.

If the collision rate is now increased from zero, the eigenvalue increases approximately with the corresponding term in the matrix
\be
  \xi^{(2)}_3 ~\sim~ {f_1M_1\over v_s^2}\left({T_2\over T_1}\right)^{-3/2} ~~.
\ee
This is the spatial rate at which the electron and ion temperatures exponentially approach one another as they converge to their downstream value.  Its inverse is the size of the ion cooling region $L_{\rm IC} \sim M_1^4/f_1$, using the hypersonic ($M_1\gg1$) limit of \eqs\ (\ref{eq:vrat}) and (\ref{eq:Trat}).

\begin{figure}[htb]
\epsscale{0.8}
\plotone{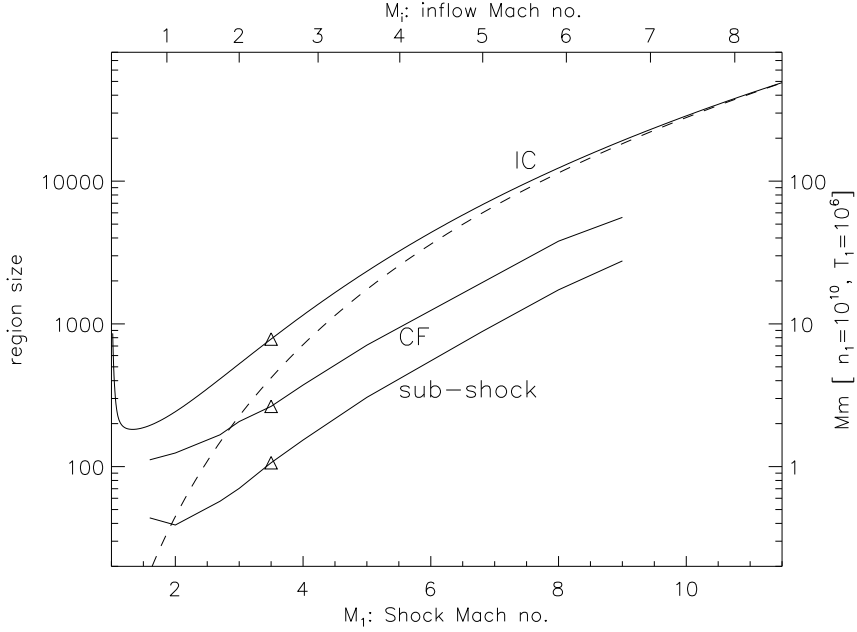}
\caption{The scales of the three regions in the stationary shock, versus the shock Mach number $M_1$ (bottom axis) and inflow Mach number $M_i$ (top axis).  The right axis are the lengths in Mm when $n_1=10^{10}\,{\rm cm^{-3}}$ and $T_1=10^6$ K.  Solid curves 
are, from top to bottom,  the lengths of the ion cooling region ({\sf IC}), the conduction front ({\sf CF}) and the sub-shock.  Triangles are the values from \fig\ \ref{fig:phase} at $M_1=3.5$.  Dashed curve is the approximation $L_{\rm IC}=0.05M_1^4/f_1$.}
	\label{fig:size}
\end{figure}

The other regions in the profile can be analyzed only using the full non-linear solution.  We define the size of the sub-shock to be twice the separation between the velocity mid-point, $v=(v_1+v_2)/2$, and the velocity 95\% to the downstream value.  The extent of the cooling region is defined to be $3/\xi^{(2)}_3$.  All lengths vary significantly with Mach number as shown in \fig\ \ref{fig:size}.  It is evident that the largest of the three regions is the ion cooling region.  This is a consequence of $\nu_{ei}$ being the smallest collision frequency.  The sub-shock is consistently the smallest of the regions, but is greater than the ambient length, 
$\ell_{i1}$ (i.e.\ unity), due to both the increased viscosity in the hotter plasma and the effects of ion conductivity.

\subsection{Fluid approximation of electron heat flux}

\begin{figure}[htb]
\epsscale{0.8}
\plotone{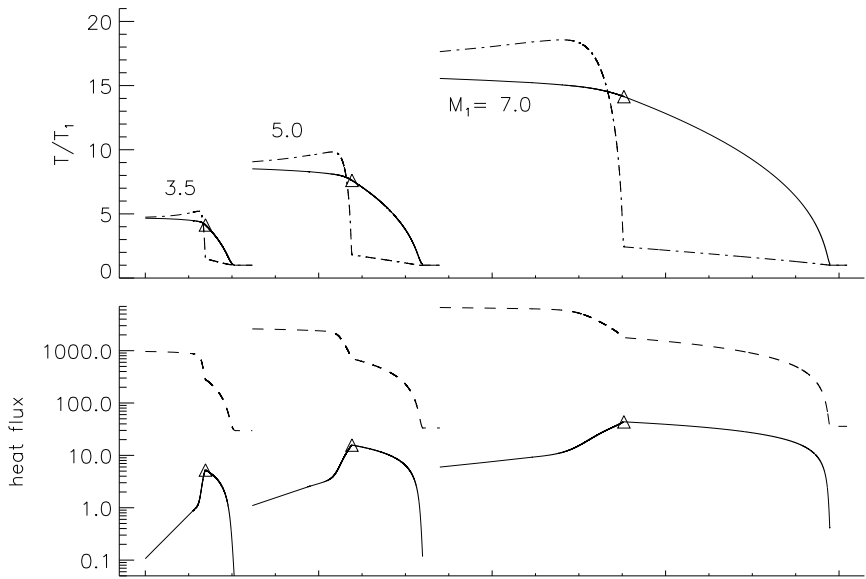}
\caption{The heat flux in stationary shocks of different Mach number.  The top panel shows temperature profiles of the electrons (solid) and ions (broken line) for $M_1=3.5$ (left), $5.0$ (middle) and $7.0$ (right).  The horizontal scale is the same for all three.  Below these are the electron heat flux $F_e$ (solid) and the free-streaming limit, $F_e^{\rm (fs)}$ (dashed) for each case.  Triangles mark the location of peak heat flux.}
	\label{fig:fe}
\end{figure}

The magnitude of the electron heat flux $-\kappa_e\nabla T_e$ is given by the dimensionless variable $F_e$ times $m_in_1c_{s,1}^3$.  This classical, local form of heat flux is derived assuming small departures from a Maxwellian electron distribution.  This approximation is expected to break down as the heat flux approaches a ``free-streaming'' value 
\citep{Campbell1984}
\be
  F_{e}^{\rm (fs)}~=~\thh m_e n\, v_{{\rm th},e}^3 
  ~=~ \thh \gamma^{-3/2}\,\sqrt{m_i/m_e}\,
  \,[m_in_1c_{s,1}^3]\Bigl\{n(T_e/T_1)^{3/2}\Bigr\} ~~,
  	\label{eq:Fefs}
\ee
where the quantities inside curly braces are the dimensionless versions.  Since both quantities scale identically with density and ambient sound speed (the factor in square brackets), their ratio will depend only on the rescaled solutions above.  Figure \ref{fig:fe} shows that the actual heat flux, $F_e$, is well below the free-streaming limit throughout the shock; in all cases shown the peak heat flux is below the limit by a factor less than 2.5\%.  We thus conclude that a classical, local heat flux expression is justified in stationary slow mode shocks and their CFs.

The fact that $F_e\ll F^{\rm (fs)}_e$ is equivalent to the statement that the temperature gradient scale, $\ell_T=|\partial\ln T_e/\partial s|^{-1}$, is much greater than the local electron mean free path, $\ell_e$, throughout the shock.  The ratio of the polarization electric field to the Dreicer electric field, the field at which thermal electrons would ``run away'', involves a similar length scale comparison, but with the pressure gradient scale in place of $\ell_T$.  This is satisfied to the same degree as the heat flux, so the polarization electric field is everywhere much less than the Dreicer field.  These statements are not entirely trivial since the gradient scales in the solution are set by diffusive processes and can therefore have length scales similar to the mean free paths which define those processes.  

\subsection{The Single-fluid limit}

Stationary versions of the single-fluid equations can be obtained by setting $T_e=T_i=T$ in \eqs\ (\ref{eq:vs}) and (\ref{eq:Tis}), and using \eq\ (\ref{eq:Tes}) to replace 
$F_e=\chi(dT/ds)/\pr_e$.  The coupled equations, after some manipulation, are
\begin{eqnarray}
  {dv\over ds} &=& 
  -{2M_1\over \chi  v}\left[ \,T - g(v)\,\right] ~~,\label{eq:v1fls} \\
  {dT\over ds}   &=& -{3M_1\pr\over \chi}\left[ T - h(v)  \right] ~~,\label{eq:T1fls}
\end{eqnarray}
with the single-fluid Prandtl number given by the harmonic sum of the electron and ion version $1/\pr=1/\pr_e+1/\pr_i=1/0.012$.

It is evident that these equations have fixed points where $T=g(v)=h(v)$, the same as the two-fluid case.  The nonlinear system is easier to solve since the upstream fixed point has two negative eigenvalues and the downstream fixed point has one positive and one negative eigenvalue.  One integrates away from the downstream fixed point along the direction of the positive's eigenvector (actually anything nearby will converge to that direction).  The solution then converges automatically to the upstream fixed point.  Such an analysis has been presented several times before in neutral gases \citep{Thomas1944,Grad1951}, governed by hard-sphere collisions where $\chi(T)=\sqrt{T/T_1}$.  \citet{Kennel1988} and \citet{Guidoni2010} used 
$\chi(T)=(T/T_1)^{5/2}$ appropriate to a high-temprature plasma.  Such a 
solution is shown along  with the two-fluid solutions in \fig\ \ref{fig:phase}.

The single-fluid shock naturally lacks an ion cooling region since the ions and electrons have the same temperature by {\em fiat}.  There is a sub-shock, where the flow decelerates, and a CF ahead of it.  The phase space curve is roughly horizontal (isothermal) between fixed point {\sf 2} and the limiting curve $T=g(v)$; this is the sub-shock.  The phase-space curve then approaches fixed point {\sf 1} following $T\simeq g(v)$ --- the conduction front.  

This simplified two-part structure cannot occur when $v_2$ is to the right of the peak in $g(v)$.  Setting these two equal gives an equation 
\be
  v_2 - v_{\rm peak} ~=~ 0 ~=~ -\quarter M_{1} - \hbox{${3\over4}$}{1\over M_{1}}
  - \left( -\half M_{1} - \hbox{${3\over10}$}{1\over M_{1}} \right)  ~~,
\ee
whose solution is the critical Mach number $M_{{\rm cr},1}=3/\sqrt{5}=1.34$, below which no CF occurs.  The same curve, $g(v)$ governs the two-fluid case, so we expect a critical transition there too.  The critical shock Mach number corresponds to a very small inflow Mach number $M_{{\rm cr},i}=0.44$, expected only after reconnection at very small angles: for $\beta_0=0.01$, shown in \fig\ \ref{fig:brat}, sub-critical reconnection requires 
$\Delta\theta<30^{\circ}$.


\section{Development of the shock}

We expect the post-reconnection flow to approach the stationary solutions found above.  The inflow boundary of our shock tube is intended to represent the retraction flow established in region {\sf 1} by the rotational discontinuity (RD).  By moving the inflow boundary to $s\to+\infty$ we have assumed that the RD recedes rapidly enough that it does not affect the central flow.  It will not, however, initialize a flow resembling the stationary shock.  We propose here a simple form for the initialization of our shock tube problem intended to mimic the effect of the reconnection which created the tube and the RDs which created the flow.

We neglect any heating that might have occurred during the reconnection event that created the post-reconnection flux tube.  We thereby take the thermodynamic quantities to be initially uniform: $T_e=T_i=T_1=0.3$ and $n=n_1=1$ in our rescaled variables.  This approach differs significantly from some previous investigations that focussed exclusively on the energetics of the dissipation region --- for example on Joule heating.  Our approach is, however, entirely consistent with the Petschek scenario in which most of the heating occurs outside the diffusion region in shocks.  Moreover, there is sufficient uncertainty about reconnection microphysics that no consensus exists on how any initial heat might be apportioned between electrons and ions.  Beginning with uniform ambient properties, as we do, is a conservative means of side-stepping the issue altogether.   In any event, the diffusion region is so small that any initial energy deposited will be quickly overwhelmed by the energy thermalized in shocks driven by subsequent retraction.

For numerical convenience we choose a smooth profile with which to initialize the velocity
\be
  v(s,0) ~=~ -M_i\tanh(s/\lambda) ~~,
\ee
where the initial gradient scale, $\lambda$, represents the size of the original diffusion region.  We find that any length smaller than $\ell_{i1}$ (i.e.\ $\lambda<1$) leads to similar behavior as the viscosity quickly smooths the velocity out to this scale.  In the opposite limit, the length scale $\lambda$ defines a ballistic time scale, 
$\tau_{\rm b}=(\lambda/M_i)\tau_{i1}$ --- the time on which flow would collapse the profile to a singularity. 

\subsection{Time-dependent solutions}

We numerically solve the dynamical equations, (\ref{eq:cont}), and (\ref{eq:mom})--(\ref{eq:Ei}) subject to inflow boundary conditions at a right boundary, $s=L$, and the initial conditions described above.  We take the left boundary to be $s=0$ with mirror conditions $v(s)=0$ and $\partial n/\partial s=\partial T_e/\partial s=\partial T_i/\partial s=0$ there.  Alternatively we could have placed a second boundary at $s=-L$ from which to drive rightward inflow, $v(-L)=+M_i$.  The center would have been mirror symmetric owing to the symmetry of the inflows.

\begin{figure}[htp]
\epsscale{1.0}
\plotone{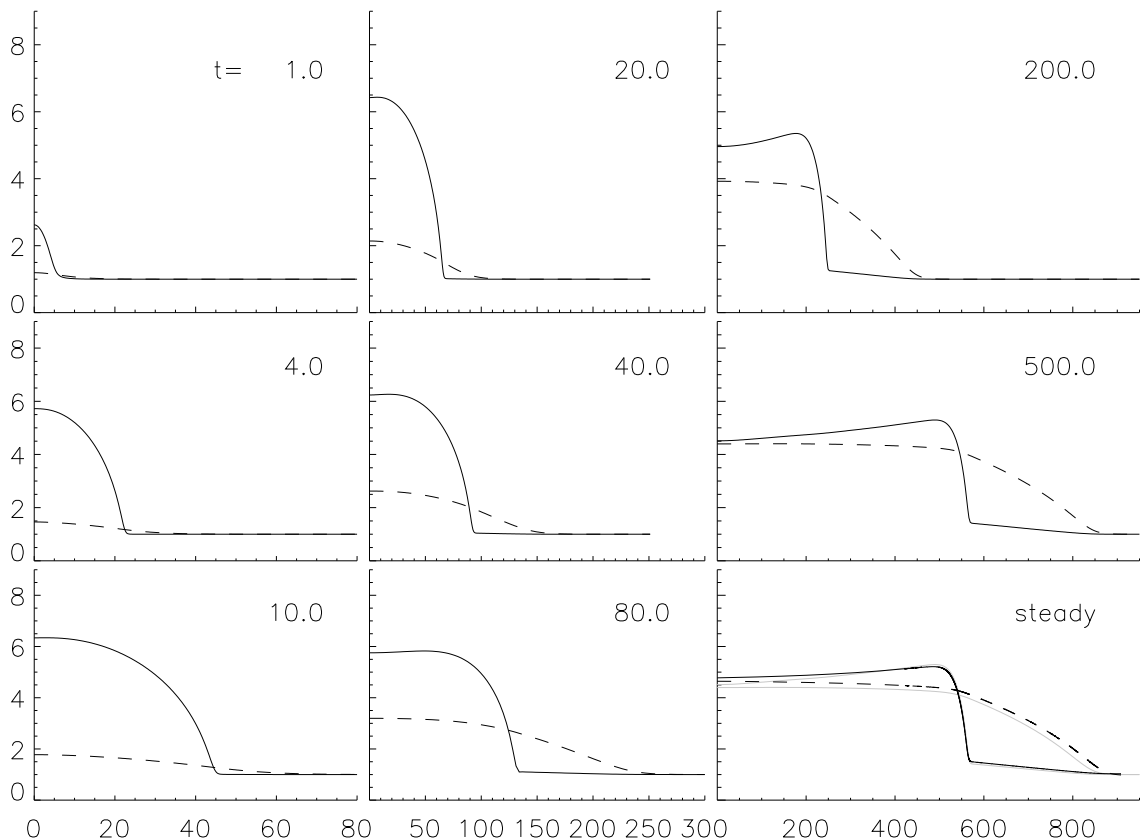}
\caption{Temperature of the ions (solid) and electrons (dashed) at times during the development of a shock.  Inflow is at $M_i=2.4$ with initial gradient scale $\lambda=5\ell_{i1}$.  The first eight panels are from successive times from 
$t=1.0\tau_{i1}$ (top left) to $t=500\tau_{i1}$ (center right).  The horizontal axis is different
in each column.  The bottom right panel is the stationary solution from \fig\ \ref{fig:phase} ($M_1=3.5$).  Grey curves are the time-dependent solution from the panel above.}
	\label{fig:smry}
\end{figure}

Figure \ref{fig:smry} shows the development of a shock from an inflow of $M_i=2.4$  beginning with a gradient scale of $\lambda=5\ell_{i1}$.  The evolution exhibited here is typical of other Mach numbers.  A region of high ion temperature develops initially and the region expands into the inflow.   The central electron temperature begins to rise slowly and gains speed once the ion temperature begins falling.  It is at this phase that the electron temperature peak spreads out, ultimately creating the CF upstream.

\begin{figure}[htp]
\epsscale{0.95}
\plotone{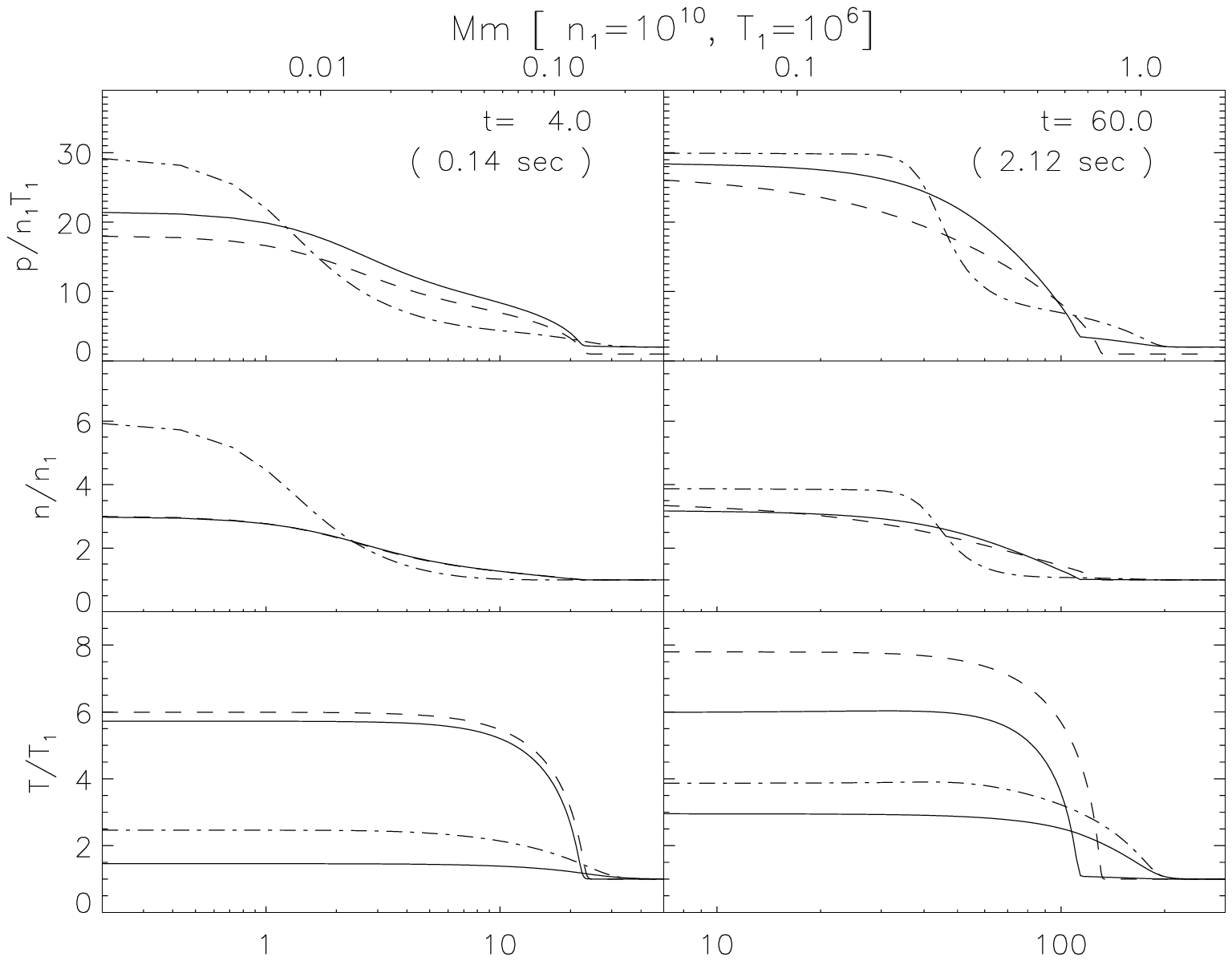}
\caption{Snapshots from the evolution of the two-fluid system (solid), the ion-fluid system (dashed) and single-fluid system (broken) at $t=4$ (left) and $t=60$ (right).  The two-fluid and single-fluid systems have $M_i=2.4$ while the ion-fluid has 
$\tilde{M}_i=\sqrt{2}M_i=3.40$.  The bottom panels show the temperature against a logarithmic length variable.  The upper and lower solid curves are the ion and electron temperatures respectively.  The center panels show the density at the same times.  The top panels show total pressure (solid), the single-fluid pressure (broken) or the ion pressure of the ion-fluid.  The top axis shows Mm for the case 
$n_{e1}=10^{10}\,{\rm cm^{-3}}$ and $T_1=10^6$ K.}
	\label{fig:ion_2fl_cmp}
\end{figure}

In the initial phase the ions are heated without appreciably affecting the electrons.  Their dynamics can be approximated by that of a single fluid with half the pressure and whose sound speed is thus $\tilde{c}_{s1}=c_{s1}/\sqrt{2}$; we call this the ion-fluid.  It is a flow with an effective Mach number  $\tilde{M}_i=\sqrt{2}M_i$ and thermal conductivity from the ions alone: $\tilde{\pr}=\pr_i=0.33$ rather than $\pr=0.012$ for the single-fluid system.  Figure \ref{fig:ion_2fl_cmp} shows the state of a two-fluid system with $M_i=2.4$ and an ion-fluid at comparable times.  Owing to its lower sound speed the  ion-fluid solution is plotted against a coordinate rescaled to 
$\tilde{\ell}_{i1}=\ell_{i1}/\sqrt{2}$.  The two-fluid system was initialized with gradient 
scale $\lambda=5\ell_{i1}$ and the ion-fluid with $\lambda=5\tilde{\ell}_{i1}$.

\begin{figure}[htp]
\epsscale{1.0}
\plotone{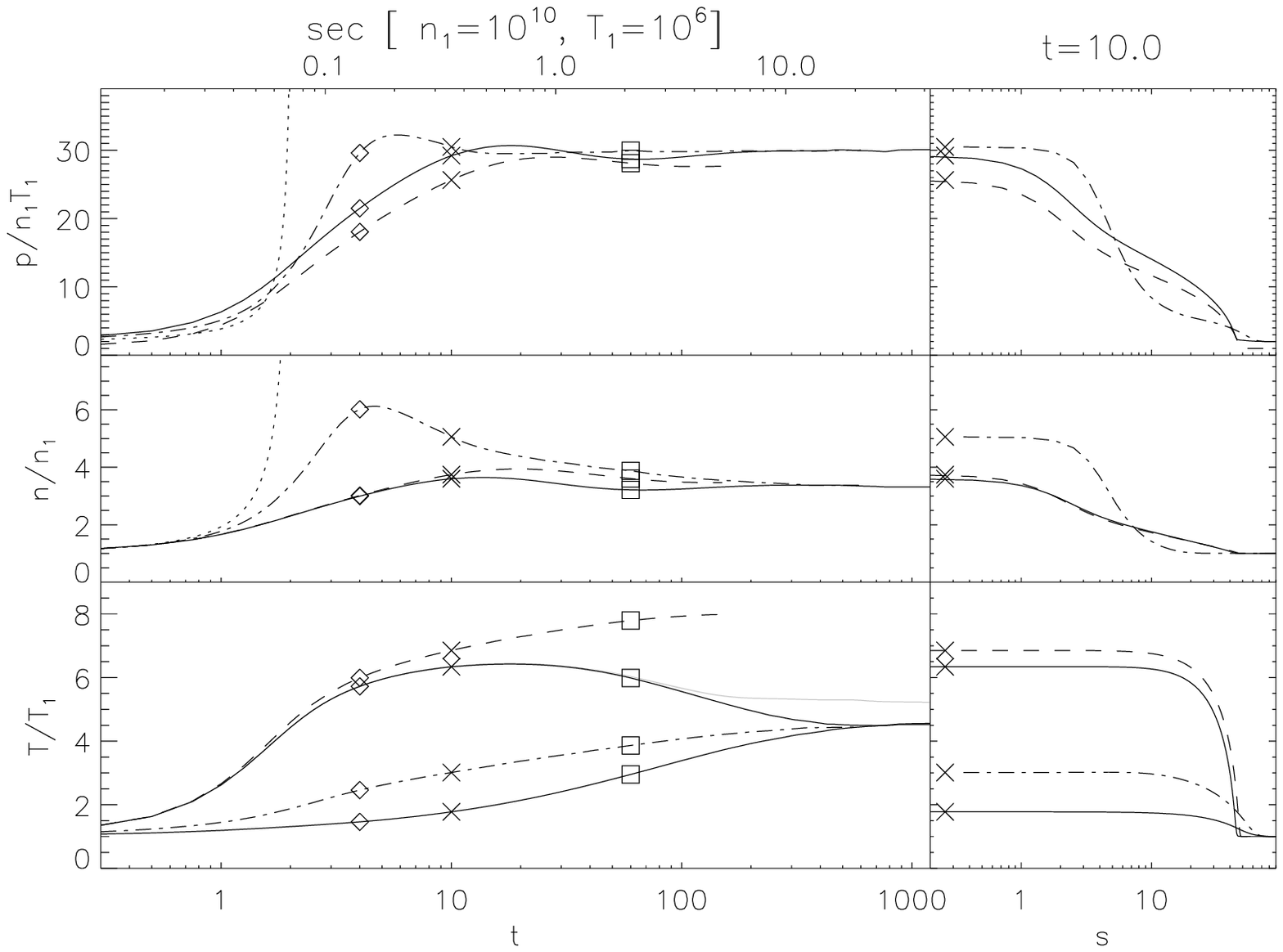}
\caption{Time history of the central values for the $M_i=2.4$ flow.  The left panels show the temperature (bottom), density (middle) and pressure (top) {\em vs.} time; the bottom axis is in rescaled units and top axis in seconds for $n_{e1}=10^{10}\,{\rm cm^{-3}}$ and $T_1=10^6$ K.   Solid curves are from the two-fluid solution; both electron temperature (lower curve) and ion temperature (upper curve) are plotted in the bottom row.  
Dashed and dash-dotted curves are the corresponding ion-fluid and single-fluid solutions. A grey line shows the peak ion temperature where that is different from the central value.   The total pressure ($p_e+p_i$) is plotted in the top row.  A dotted line in the middle and top rows show isothermal, ballistic evolution which is singular at $t=\tau_{\rm b}$.  Symbols show the values at times where profiles are plotted in \fig\ \ref{fig:ion_2fl_cmp} (diamonds and squares) or in the right panels (crosses).  The right panel show snapshots at $t=10$ in the same format as \fig\ \ref{fig:ion_2fl_cmp} as well as the peak values (crosses) which are represented to the left.}
	\label{fig:peak_thist}
\end{figure}

The spatial distributions of temperature, density and pressure are generally peaked at the center, $s=0$.  Figure \ref{fig:peak_thist} plots these central values versus time for this particular case, $M_i=2.4$.  This along with \figs\ \ref{fig:smry} and \ref{fig:ion_2fl_cmp} illustrate and explain the four basic phases which compose the shock development in the two-fluid system. Except in cases with large $\lambda$, discussed below, solutions at all Mach numbers pass through the same phases but at different times.

The first phase is ballistic collapse during which the initial velocity gradient is steepened and density and pressures begin to rise.  In the present example this occurs for 
$t\la\tau_{\rm b}=2.1$.   In the second phase ($2\la t\la10$), called {\em ion heating},  the ions are heated as they would be in a pure ion-fluid.  The ion-fluid temperature will approach the steady value corresponding to a larger Mach number, 
$\tilde{M}_1=\sqrt{2}M_1$.  This is greater than, $T_2$,  so the ion temperature overshoots its final Rankine-Hugoniot value. At very large Mach number, $\tilde{T}_2=2T_2$, so the overshoot can be considerable. The left column in \fig\ \ref{fig:smry} consists of times during ion heating. 

In the third phase, {\em front development} ($10\la t\la 100$), the ions transfer heat to the electrons who transport it via thermal conduction to a growing CF.  At the end of this phase the density and total pressure, $p_e+p_i$ have achieved their final values, but the electrons are still colder than the ions at the center.   The middle column in \fig\ \ref{fig:smry} consists of times during front development.  In the final phase ($t\ga 100$) the ion cooling region develops in the center.  The central curvature in the ion temperature reverses and the new pair of peaks move outward with the sub-shock.  The central concavity  descends to form a growing ion cooling region (see the right column in \fig\ \ref{fig:smry}).  The decreasing central ion temperature meets the increasing central electron temperature  at  a level slightly below their equilibrium down-stream value, $T_2$; they then increase together toward this final value ($t\ga500$).

The single-fluid system evolves in a notably different manner than the two fluid system.  During the ballistic collapse phase the density of all systems rise passively as
\[
  n(0,t) ~\simeq~ {1\over 1-t/\tau_{\rm b}} ~~,
\]
shown as a dotted line in \fig\ \ref{fig:peak_thist}.  This ends when pressure builds to a level sufficient to end the collapse. The collapse progresses farther in the single-fluid system because its large thermal conductivity makes it approximately isothermal, 
$p~\sim~(1-t/\tau_{\rm b})^{-1}$.  In contrast to this, collapse in the pure-ion fluid, and thus in the two-fluid system, is better approximated as adiabatic
$p~\sim~(1-t/\tau_{\rm b})^{-5/3}$.  It is for this reason that the central density in the single-fluid system overshoots the Rankine-Hugoniot value far more than in the other two systems.

Following the initial density overshoot, the central region of the single-fluid solution is forced to expand ($t\ga4$).  The expansion has a tendency to drive {\em down} the  central temperature, and thus requires additional conduction from the shock to offset it.  This competition creates a concave $T$ profile in the center, unrelated to ion cooling (obviously), which slows down the approach to final temperature.  In the $M_i=2.4$ case shown in \fig\ \ref{fig:peak_thist} the final value is not achieved until after $t=1000\tau_{i1}$.  The ion-collision time in the post-shock fluid is shorter that $\tau_{i1}$ by a factor, $(T_2/T_1)^{-3/2}=0.017$, so the approach is truly {\em slow}.

\subsection{The Ballistic collapse}

\begin{figure}[htp]
\epsscale{1.0}
\plotone{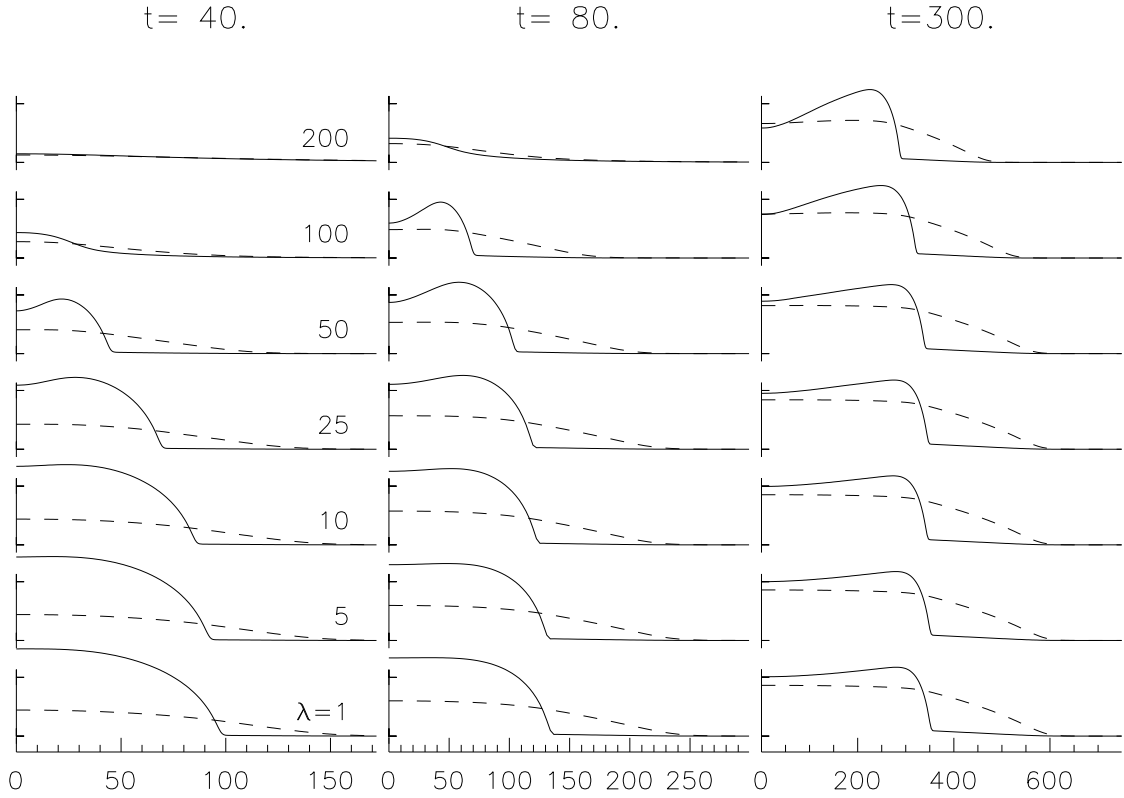}
\caption{Profiles of the ion (solid) and electron (dashed) temperatures at characteristic times for the flow with $M_i=2.4$.  Profiles are shown for flows initialized with different gradient scales, arranged in rows with $\lambda$ increasing upward from $\lambda=1$ (bottom row) to $\lambda=200$ (top row).  The vertical axis on each plot has tick marks at $T_1$ and $T_2$.}
	\label{fig:lam_scan}
\end{figure}

The scenario above is altered in cases of extended ballistic collapse.  The collapse time, $\tau_{\rm b}=\lambda/M_1$, is controlled by the initial gradient scale $\lambda$.  For very small values, $\lambda\la1$, viscosity smoothes the initial gradient very rapidly and all subsequent evolution is unchanged.  Larger values, on the other hand lead to a modified scenario illustrated by the upper curves of \fig\ \ref{fig:lam_scan}.

For values of $\lambda$ large enough, the collapse phase can overwhelm the ion heating phase with consequences visible in the upper two rows of \fig\  \ref{fig:lam_scan}.  The slow collapse drives up both ion and electron temperatures adiabatically -- they increase together.  Electron conductivity then begins to generate a CF even before the ions have been heated.  A shock ultimately forms away from the center, leaving the central ions unheated.  These then cool as they subsequently expand, driving down the central ion temperature.  This central depression superficially resembles the ion cooling region, but due to its different formation process it is a deeper concavity and the ion temperature can actually fall below the electron temperature (the upper right profile, $\lambda=200$, $t=300$).  Nevertheless, the shock conforms to the steady-state morphology and approaches the same steady profile.  Its slower formation means the shock is slightly behind the location it would have had with smaller 
$\lambda$.

\subsection{Motion of the shocks and fronts}

\begin{figure}[htb]
\epsscale{0.7}
\plotone{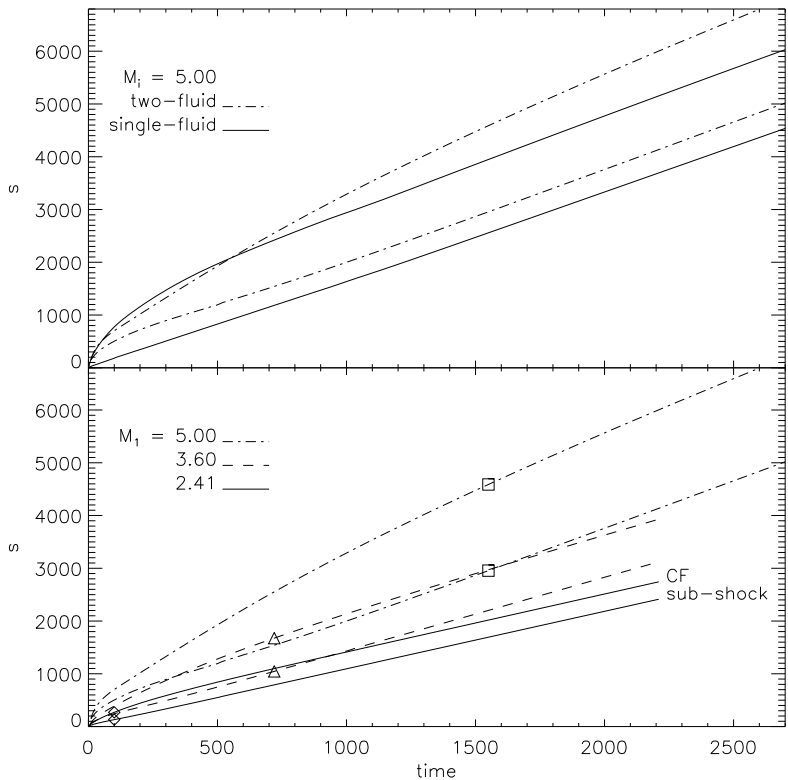}
\caption{Positions of the conduction front (CF) and sub-shock over time for shock-tube solutions.  The bottom panel compares two-fluid solutions with different Mach numbers: $M_i=2.4$ (solid), $3.6$ (dashed) and $5.0$ (broken).  For each solution the upper curve is the CF and the lower is the sub-shock.  Symbols on each pair show the beginning of the ion cooling phase. The top panel compares $M_i=5.0$ solutions of the two-fluid problem (broken) with the single-fluid solution (solid). }
	\label{fig:cf}
\end{figure}

Shock tube solutions at different Mach numbers show the same development described above, but at different times.  Figure \ref{fig:size} shows that the extents of regions in the steady-state solutions scale strongly with $M_1$, and therefore with $M_i$.  Figure \ref{fig:cf} shows how these regions move in time-dependent solutions.  Each expands from zero to its final size.  The higher Mach number cases have farther to go and thus develop more slowly.  The ion cooling region is the largest and its development begins (symbols) substantially later for larger Mach numbers.  The CF has attained its final size once its leading edge moves with the sub-shock --- i.e.\ two curves in \fig\ \ref{fig:cf} have become parallel, as they are for $t\ga800\tau_{i1}$ in the $M_i=3.6$ solution (dashed curves).

The different front development in the single-fluid system affects the motion of its CF as shown in the top panel of \fig\ \ref{fig:cf}.  Nearly instantaneous diffusion steals some of the post-shock pressure from the single-fluid sub-shock, causing it to lag behind its two-fluid counterpart.  While the final shock speeds are identical for the two cases (from the Rankine-Hugoniot conditions) its higher pressure early on gives the two-fluid case a lead it never loses.  The same instantaneous diffusion in the single-fluid system propels its CF  ahead of its two-fluid counterpart at early times.  This development appears more clearly diffusive, $s_{CF}\sim \sqrt{t}$, than the two-fluid does.  The steady-state CF width is, however, greater in the two-fluid case (see e.g.\ \fig\ \ref{fig:phase}), so it eventually regains the lead.   For the $M_i=5$ case in the figure, it does this at $t\simeq500\tau_{i1}$.  Finally, it is clear that front development lasts far longer in the two-fluid case: the single-fluid curves 
are parallel at a much earlier point ($t\simeq800\tau_{i1}$) indicating the CF has attained its final width.

\subsection{Validity of the fluid approximation}

During the evolution toward steady state, the flow develops gradients whose scale may challenge the assumptions underlying fluid approximation.  For example, gradient length scales shorter than the mean free path violate the Chapman-Enskog derivation of classical, local heat flux.  We demonstrated above that, at least in steady state, the electron temperature gradients are gradual enough to justify this approximation.  We find the same to be true during the time-evolution.  This finding contradicts the hypotheses of many previous investigations \citep{Brown1979,Smith1979,Karpen1987} which proposed modified treatments of the electron heat flux to handle the strong heating expected in flares.  Our findings are different because, while those investigators assumed electrons were heated directly, our electrons are heated through collisions with a much hotter ion population, which had themselves been heated in the shock.  Since that collisional coupling is far weaker than the conductivity, strong gradients never develop in the electron temperature.

Since it is the ions that are first heated it is their temperature gradient that is most likely to violate the fluid approximation.  Indeed, we find that $F_i/F_i^{\rm (fs)}$ becomes comparable to, and sometimes slightly greater than, unity during early phases of shock development.  This is symptomatic of a more serious issue involving the viscosity itself.  

Viscosity is the ultimate source of heat but also contributes to decelerating the flow through the viscous stress.  The pressure combines with the parallel component of the viscous stress tensor, $\sigma_{ss}$, to produce a diagonal element of the pressure tensor 
\citep{Lifshitz1981}
\be
  p_{ss} ~=~ p_i - \sigma_{ss} ~=~ p_i - \hbox{${4\over3}$}\mu_i{\partial v\over\partial s} ~~.
\ee
This is a second moment of the ion distribution function which can never be negative.  This means that the velocity gradient must satisfy the condition
\be
  \hbox{${4\over3}$}{\mu_i\over p_i}{\partial v\over\partial s} ~=~ 
  \chi_i{\partial v\over\partial s} ~\le~1 ~~,
  	\label{eq:visc_cond1}
\ee
where the second expression involves rescaled variables.  

The velocity profile in a shock is principally compressive, $\partial v/\partial s<0$, so condition (\ref{eq:visc_cond1}) is always satisfied.  Even though we need not consider perpendicular momentum in our one-dimensional shock tube problem, there are diagonal perpendicular terms in the pressure tensor which must also be non-negative.  The terms in the viscous stress, $-\sigma_{ss}/2$, will be positive in the case of compression and partially cancel out the pressure.  Demanding non-negativity of this term completes the condition on the velocity gradient
\be
  -2 ~\le~ \chi_i{\partial v\over\partial s} ~\le~ 1 ~~,
  	\label{eq:visc_lim}
\ee
where the upper bound is \eq\ (\ref{eq:visc_cond1}).

Violation of criterion (\ref{eq:visc_lim}) means one of the velocity second moments is negative.  Since the squared velocity can never be negative this indicates that the distribution function itself is negative over a significant region in velocity space.  While this is clearly unphysical, it is a common result of applying the Chapman-Enskog procedure beyond its range of validity \citep{Campbell1984}: the equilibrium Maxwellian has been supplemented by a negative perturbation so large the sum itself becomes negative.  Nor does a positive value of the second moment prove the distribution function was everywhere positive.  Only for 
$\chi_i|\partial v/\partial s|\ll1$ can we be sure this is true, and that the classical, local transport formulae apply.

\begin{figure}[htp]
\epsscale{0.9}
\plotone{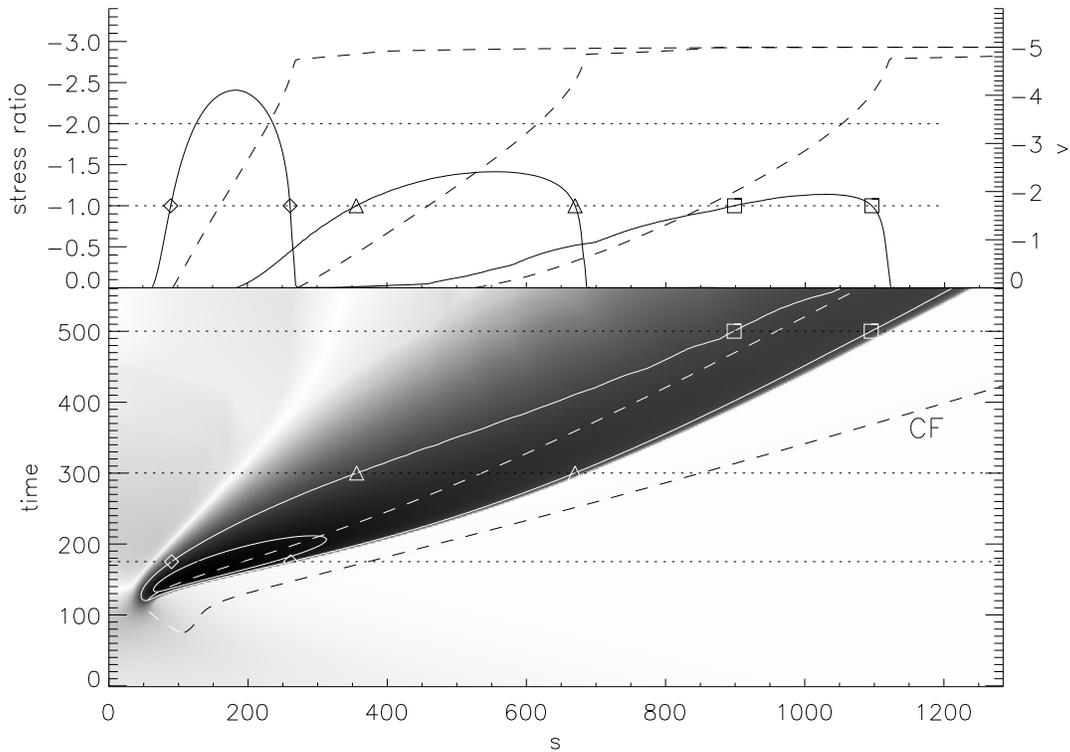}
\caption{A plot of the rescaled viscous stress, $\chi_i\partial v/\partial s$, during the time development of a shock from $M_i=5$ and $\lambda=500\ell_{i1}$.  The quantity is plotted as a logarithmic grey-scale in the bottom panel against time (vertical axis) and position (horizontal axis).  Solid contours are for values of $-1$ (outer) and $-2$ (inner).  Dashed curves are the locations of the sub-shock (upper) and CF (lower).  Horizontal dotted lines indicate three different times at which the viscous stress is plotted in the top panel (solid curves).  Symbols show where the value crosses $-1$ in each slice.  The velocity itself is shown as a dashed curve at all three times, plotted against the right axis.  Note that both axes are negative going upward, since both viscous stress and velocity are negative.}
	\label{fig:visc}
\end{figure}

Figure \ref{fig:visc} shows $\chi_i\partial v/\partial s$ for all times and positions from a solution with $M_i=5$ and $\lambda=500\ell_{i1}$.  It is small for $t<\tau_{\rm b}=\lambda/M_i=100$, and soon therafter violates criterion (\ref{eq:visc_lim}).  The violation is short-lived and confined to $s<300$, but the region in the vicinity of the sub-shock continues to harbor values above half the limit; classical viscosity is probably inaccurate there.  Outside the sub-shock itself, however, the condition is well satisfied and the fluid solution is probably valid.

Shock solutions using methods beyond the fluid closures find very similar structure with slightly thicker sub-shock (ion shock) embedded inside the CF and ion cooling region.  Steady state solutions have been found with the Fokker-Planck equation in the steady-state method of \citet{Mott-Smith1951} \citep{Greywall1975,Abe1975}.  Time-dependent solutions have been found using fully kinetic treatment of ions along with fluid electrons \citep{Casanova1991,Vidal1993}.  All such treatments find solutions very similar at modest Mach numbers ($M_i\simeq2$).  Solutions at larger Mach numbers ($M_i\ga 5$) have sub-shocks 2--4 times thicker than fluid solutions, but are otherwise extremely similar.  We thus consider our fluid solutions to be reasonable representations of the actual behavior in post-reconnection outflows.

\section{The Slow shocks and conduction fronts in a reconnection jet}

The time-dependent shock solutions can be placed back into the context of Petschek reconnection to exhibit the structure of an outflow jet .  While the analysis has been time-dependent, we place the time-evolving solutions into a traditional  steady-state model viewed from the end-on perspective of 
\fig~\ref{fig:toon}.  

\subsection{The jet geometry}

The configuration is structured by the rotational discontinuity (RD) which propagates along the inner field line, against the inflow, at the Alfv\'en speed.  Its location in the rescaled shock-tube variables, $s$ and $t$,  is
\be
  s_{RD} ~=~ \left( {v_{{\rm A}0}\over c_{s1}} - M_i \right)\, t ~=~
  \left( \sqrt{1.2\over \beta_0} - M_i \right)\, t 
  ~=~\sqrt{1.2\over \beta_0}\cos(\Delta\theta/2)\,t ~~,
  	\label{eq:sRD}
\ee
where $\Delta\theta$ is the angle between field lines across the initial current sheet (see \fig\ \ref{fig:toon}) and the final expression uses \eq\ (\ref{eq:Mi}) for $M_i$ and 
takes $\Omega_{0,1}=\Delta\theta/2$. Until now we have assumed $\beta_0$ to be small enough that $s_{RD}$ was effectively infinite, or at least was ahead of the CF.  

Inside the RDs ($s<s_{RD}$) field lines have been advected downward at the vertical Alfv\'en velocity, $v_{{\rm A}0y}=v_{{\rm A}0}\sin(\Delta\theta/2)$.  They have been advected for $\Delta t$ since reconnection at the X-point, giving them a vertical position
\be
  y ~=~ v_{{\rm A}0y}\Delta t ~=~ v_{{\rm A}0}\sin(\Delta\theta/2)\,{\ell_{i1}\over c_{s1}}t ~=~
  \sqrt{1.2\over\beta_0}\, \sin(\Delta\theta/2)\,\ell_{i1}\,t ~~,
  	\label{eq:y_pos}
\ee
where $t$ is the re-scaled time since reconnection ($t=0$) and $y$ increases downward.  For simplicity, we assume that the X-point is stationary, but it would be easy to add to 
\eq\ (\ref{eq:y_pos}) an arbitrary X-point motion.

The relation between the shock-tube coordinate $s$ and the horizontal $x$  coordinate normal to the current sheet depends on aspects of reconnection we have not had to consider until now.  The magnetic field {\em normal} to the current sheet, $B_{x0}$,  depends on the external structure in steady state models \citep{Petschek1964} or the fast magnetosonic rarefaction wave in transient models \citep{Lin1993,Nitta2002}.  It depends on the reconnection rate (heretofore unspecified) and thereby determines the angle 
$\alpha=\hbox{atan}(B_{x0}/B_{y0})$ the field lines make with the vertical in the end-on view.  External solutions typically find this to be of order $10^{\circ}$ or less.  The value has no physical significance for our calculation so we adopt $\alpha=10^{\circ}$ for clarity of illustration. 

The field lines inside the RDs are horizontal and their length is parameterized by our rescaled shock-tube coordinate $s$.  These horizontal field lines, with strength $B_1$, make an angle 
$\hbox{asin}(B_{x1}/B_1)$, with the line-of-sight (the $z$ axis).  The horizontal coordinate in the end-on view is therefore
\be
  x ~=~ {B_{x1}\over B_1}\ell_{i1}\, s ~=~ {B_{x1}\over B_{y0}}\,{B_{y0}\over B_1}
  \ell_{i1}\, s ~=~
  \tan\alpha\,\sin(\Delta\theta/2)\,\ell_{i1}\, s ~~,
  	\label{eq:x_pos}
\ee
where the final expression uses the fact that $B_1=B_0$ and $B_{x1}\simeq B_{x0}$.  Combining expressions 
(\ref{eq:sRD})--(\ref{eq:x_pos}) gives the half-angle the RDs make in the end-on view
\be
  \tan\varphi_{RD} ~=~\left.{x\over y}\right\vert_{RD} ~=~
  \tan\alpha\cos(\Delta\theta/2) ~~,
\ee
which is naturally smaller than $\alpha$ since the external field lines must intersect the RD.

\begin{figure}[htp]
\epsscale{0.75}
\plotone{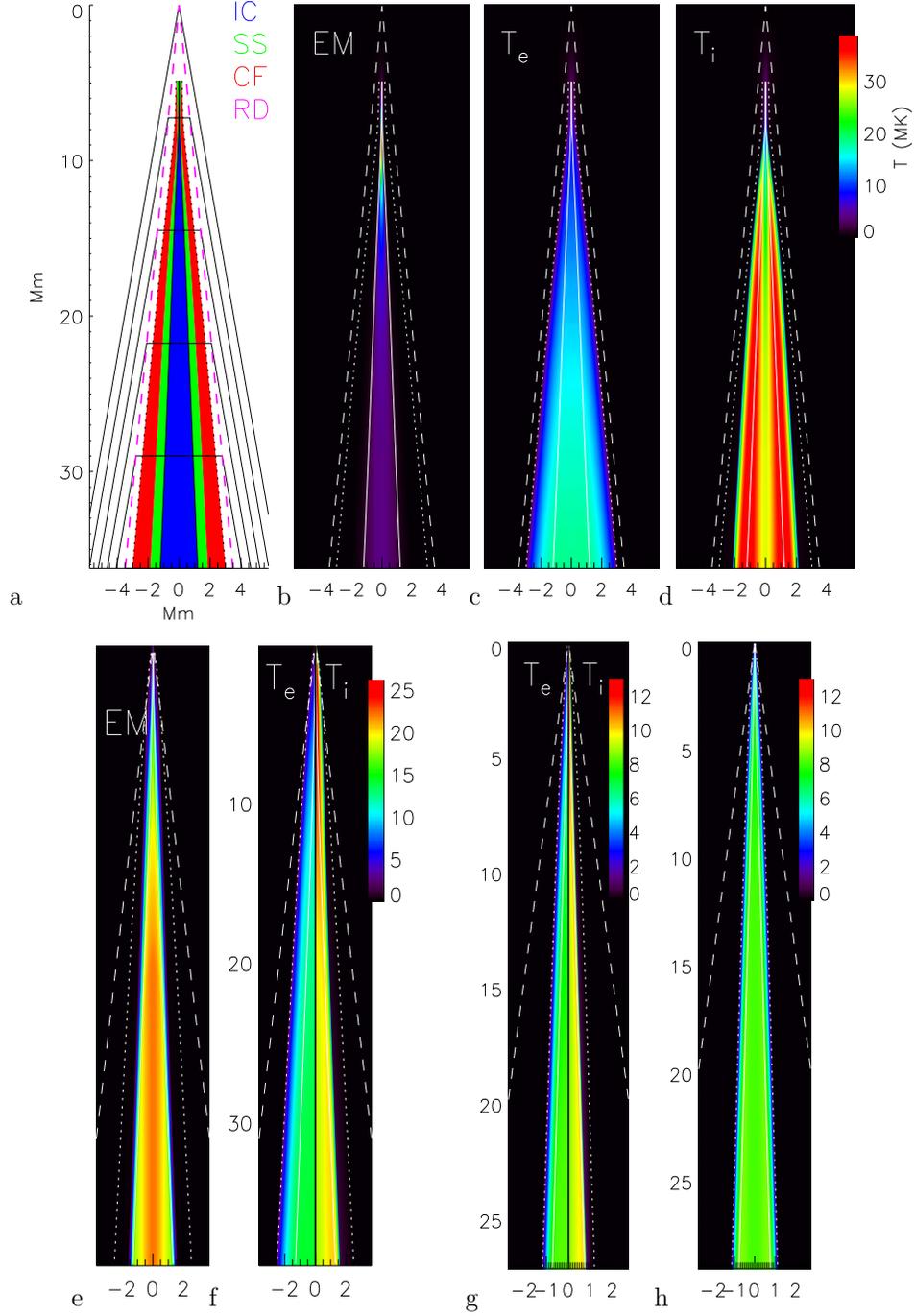}
\caption{Outflow jets synthesized from shock-tube solutions at different conditions.  
Top row:
$\Delta\theta=114^{\circ}$ ($M_i=5$) showing (a) the regions, (b) 
the emission measure, (c) $T_e$ and (d) $T_i$.  Bottom row
$\Delta\theta=95^{\circ}$ ($M_i=3.6$) showing (e) emission measure and (f) a combination of temperatures.  Panels (g--h) are for $\Delta\theta=77^{\circ}$ ($M_i=2.4$), from (g) two-fluid
and (h) single-fluid solutions.}
	\label{fig:jets}
\end{figure}

To cast the reconnection scenario in context we use the following values for the pre-reconnection plasma: $n_{e0}=2\times10^{10}\, {\rm cm}^{-3}$, 
$T_{0}=2\times 10^6$ K and $B_0=10$ G.  These choices dictate the parameters used in the rescaling, $\beta_0=0.028$, $\ell_{i1}=11.7$ km and $\tau_{i1}=0.049$ sec.   We set the external field angle at $\alpha=10^{\circ}$ and choose the current sheet angle, $\Delta\theta$, corresponding to a particular shock-tube Mach number;  for example, to obtain $M_i=5.0$ we use $\Delta\theta=114^{\circ}$.   Mapping the time-dependent $M_i=5$ solution, with 
$\lambda=500\ell_{i1}=5.9$ Mm, onto the reconnection outflow geometry yields the emission measure and individual temperature shown along the top row of \fig\ \ref{fig:jets}
(\ref{fig:jets}a--d).

The outflow jets differ in predictable ways from solutions of the full MHD equations previously reported \citep{Forbes1983,Yokoyama1997}.  The emission measure (EM) is enhanced only within the center-most region confined by the sub-shocks.  There are temperature enhancements outside this, representing the CFs.  The profile of $T_e$ extends the farthest, but $T_i$ reaches the greatest value.  The overshoot of $T_i$ creates high-temperature ridges following the sub-shocks, between which is a cooler channel from the ion-cooling region.  Both EM and $T_i$ are most enhanced within a short narrow region close to the X-point.  This is a result of ballistic collapse due to the very large initial gradient-scale, $\lambda=500\ell_{i1}$.  All of these regions are, however, inside the RDs and therefore inside the outflow jet itself (dashed lines).  This is the most notable difference from the simulations which were done for strictly anti-parallel reconnection, 
$\Delta\theta=180^{\circ}$.

The bottom row in \fig\ \ref{fig:jets} shows how the jets change at different Mach numbers and gradient scales.  All use gradient $\lambda=10\ell_{i1}=0.1$ Mm, much smaller than the top row.  As a result the sub-shocks connect to the X-point and the EM lacks an upper enhanced region.  Instead the EM increases monotonically away from the X-point.  The $T_i$ ridge is progressively muted and the IC becomes progressively narrower moving from the largest reconnection angle ($\Delta\theta=114^{\circ}$, \fig\ \ref{fig:jets}d) to the smallest 
($\Delta\theta=77^{\circ}$, \fig\ \ref{fig:jets}g).  At angles below this the region is so narrow that no structure is visible unless the horizontal axis is exaggerated --- hence we do not show them.

The two panels at the lower right (\ref{fig:jets}g--h) show the difference between electron temperature in a two-fluid (\ref{fig:jets}g) and single-fluid solution (\ref{fig:jets}h).  The main difference is that the slower rise in the former displaces the high-temperature region from X-point.  Since imaging observations generally use radiative signatures sensitive to $T_e$ (line or continuum emission, for example), they would find a high-temperature ``jet'' separated from the X-point by distances of 8--15 Mm (for $M_i=2.4$ and $3.6$ respectively).  Single-fluid models fail to predict this since the electron temperature follows the ion temperature which rises much earlier.

\subsection{Extent of the conduction front}

A significant new feature of our solutions is that the conduction fronts all lie inside the RDs which define the outflow jet.  This is notably different from  simulations of anti-parallel reconnection, \citet{Yokoyama1997} in particular, where CFs extend outside the jets where they can  drive evaporation.  Our shock-tube treatment assumed, to the contrary, that all effects from the slow shock, including the CF, remained {\em inside} the RD.  Our placement of the inflow boundary (the RD) at $s\to+\infty$ means that the shock-tube solution cannot itself violate the condition.  It is, however, possible for the assumption to be violated once we commit to a particular choice of $\beta_0$ in order to place the solution in a reconnection jet.

\begin{figure}[htp]
\epsscale{0.9}
\plotone{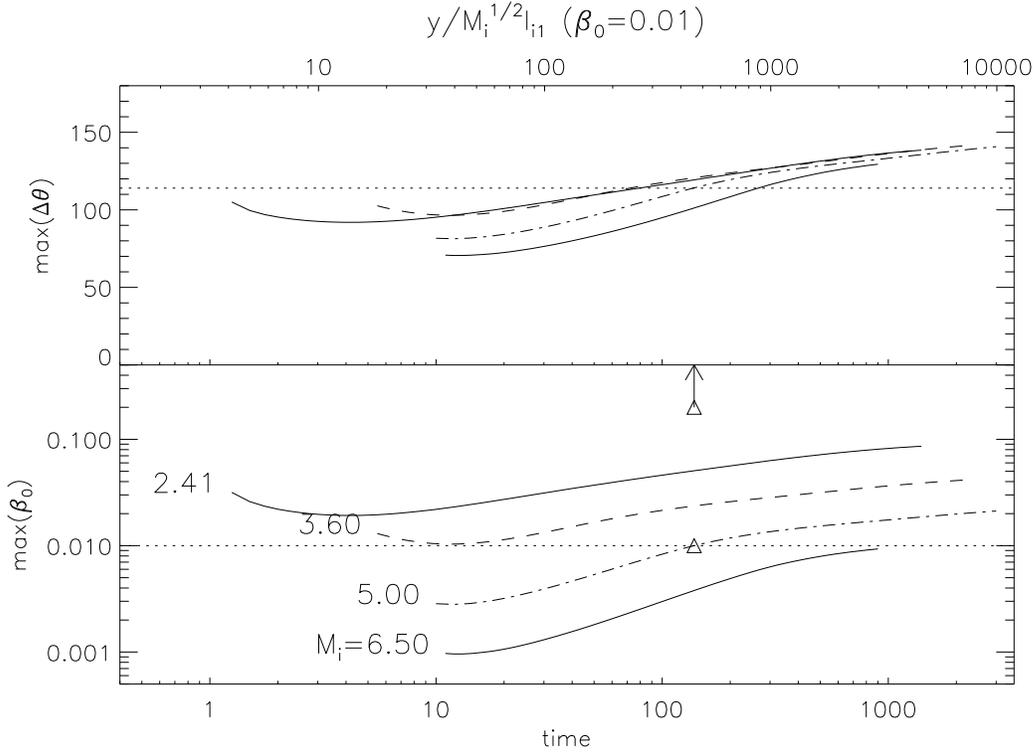}
\caption{The reconnection angle, ${\rm max}(\Delta\theta)$ (top panel) or initial beta (bottom panel) at which the CF reaches the RD for shock-tube solutions at different Mach numbers.  Expressions (\ref{eq:beta_CF}) and (\ref{eq:theta_CF}) are plotted against time (bottom axis) or against vertical coordinate in a steady-jet (top axis).  Triangle marks a case with $M_i=5.0$ discussed in the text.}
	\label{fig:max_beta}
\end{figure}

Expression (\ref{eq:sRD}), giving the shock-tube coordinate of the RD, can be solved for 
$\beta_0$ to yield
\be
  \beta_0 ~=~ 1.2\left( {s_{RD}\over t} + M_i \right)^{-2} ~~.
  	\label{eq:beta_RD}
\ee
For a shock-tube solution at a given $M_i$ we can find the position, $s_{CF}(t)$, of the CF over time, as shown in \fig\ \ref{fig:cf}.  Setting $s_{RD}=s_{CF}$ gives the condition for marginal validity of our shock-tube solutions.  Using this in \eq\ 
(\ref{eq:beta_RD}) yields the maximum value of $\beta_0$ for which a particular solution satisfies the condition
\be
  {\rm max}(\beta_0) ~=~ 1.2\left( {s_{CF}(t)\over t} + M_i \right)^{-2} ~~.
  	\label{eq:beta_CF}
\ee
Examples of this function are plotted on \fig\ \ref{fig:max_beta} for solutions at different $M_i$ and $\lambda=10\ell_{i1}$.   

The shock-tube solutions are performed at specified inflow Mach number $M_i$.  In this case the value of $\beta_0$ corresponds to a reconnection angle through \eq\ (\ref{eq:Mi}).  Using this in \eq\ (\ref{eq:beta_CF}) gives a maximum angle
\be
  {\rm max}(\Delta\theta) ~=~ 4\sin^{-1}\left[{M_i^2\over4.8}\,
  {\rm max}(\beta_0)\,\right]^{1/4}  ~~,
  \label{eq:theta_CF}
\ee
to which the internal CF assumption applies.  The top panel of \fig\ \ref{fig:max_beta} show the maximum angle for the different solutions.  This shows that for $\beta_0\la10^{-3}$ any reconnection angle $\Delta\theta\la120^{\circ}$ will have CFs inside the RD, over most of the outflow jet.

Since a CF decelerates over time, each of the curves in \fig\ \ref{fig:max_beta} trends upward.  Time in the shock-tube solution is equivalent to vertical position in the outflow jet, according to \eq\ (\ref{eq:y_pos}).  It is thus possible for the CF to reach outside the RD in the upper portion of the jet (near the X-point) and return to the inside for the remainder of the jet.  The section where $s_{CF}>s_{RD}$ would involve a pre-heating of the material ahead of the RD whose effects we have not accounted for.  For brief excursion we expect these effects to be minor since the propagation of the RD depends on density, which is far less affected by the CF than is temperature.  Thus we can use the curves in \fig\ \ref{fig:max_beta} to determine, at least approximately, the distance over which the CF extends outside the jet.

For example, reconnection at $\beta_0=0.01$ and an angle $\Delta\theta=114^{\circ}$ (dotted lines) produces an inflow of $M_i=5$.  The CF position is outside the RD for $t<120$ (triangle), corresponding to vertical coordinates $y<400\sqrt{5}\ell_{i1}$.  For $T_0=10^6$ and $n_e=10^{10}\,{\rm cm}^{-3}$, the CF extends outside the outflow over the first $5.2$ Mm of the jet; beyond that position the CF lies inside the outflow jet.  This does not occur in \fig\ \ref{fig:jets}a because that uses a solution with a large ballistic collapse, 
$\lambda=500\ell_{i1}$, for which the CF tends to lag.

Other scenarios all follow the same general pattern.  For a given reconnection angle, 
$\Delta\theta$, lower values of $\beta_0$ (i.e.\ larger $M_i$) will have larger CFs which extend outside the RD over a larger portion of the jet.  Alternatively, for a given value of 
$\beta_0$ larger reconnection angles will have more extensive CFs outside the RDs for more of the jet.  The largest angle, $\Delta\theta=180^{\circ}$, produces a CF which is everywhere outside the jet, as has been previously reported for anti-parallel reconnection 
\citep{Forbes1989,Yokoyama1997}.  (This limiting case cannot actually be treated using our shock-tube solutions.)

\section{Discussion}

We have studied the structure of the slow magnetosonic shock in the outflow from Petschek reconnection.  We have done this using two-fluid hydrodynamic equations, without including magnetic fields.  This approximate treatment is possible for cases of reconnection between sufficiently skewed fields, $\Delta\theta\la150^{\circ}$.  Such cases have been subject to less study than the special case of anti-parallel field, $\Delta\theta=180^{\circ}$.   In skewed cases the primary role of the magnetic field is plasma acceleration at rotational discontinuities (RDs).  This converts magnetic energy to bulk kinetic energy but without any heating.  The heating occurs at a slow shock downstream of the RD; its primary source is the bulk kinetic energy from the accelerated flow.   It is a nearly parallel shock which we have treated as a purely hydrodynamic shock.

Parallel shocks in ionized plasmas are best studied using two-fluid equations.  The densities and velocities of the electron and ion fluids are strongly coupled through electrostatic interactions, but their temperatures are coupled weakly though very slow collisions.  This results in a structure more complex than what would be found from a single-fluid treatment, such as MHD, wherein electron and ion temperatures are assumed equal.  The most significant feature of the two-fluid treatment is an extensive ion cooling region downstream of the ion deceleration layer (sub-shock).  The slow development of the region, on the ion-electron collision time scale, will be reflected in the region of hot electrons within the outflow.  Since most flare signatures, such as bremsstrahlung radiation, reflect the electron temperature, observations are better interpreted using two-fluid solutions than single-fluid (MHD) results.  We have exhibited several examples of outflow jets synthesized from two-fluid hydrodynamic solutions.

The most significant difference between our solution and previous studies comes from the skewed fields rather than from the two-fluid treatment.  The separation between RD and slow shock, which occurs in all skewed fields, means the conduction front is often located inside the outflow jet.  In this case the chromospheric evaporation will not precede the reconnection, as it does in the anti-parallel cases previously studied 
\citep{Forbes1989,Yokoyama1997,Seaton2009}.  Evaporation has, however, been postulated to enhance the emission measure of high temperature material to levels actually observed in flares \citep{Tsuneta1996}.  We conclude that this assertion should be made in conjunction with an estimate of the strength of the current-aligned field, sometimes called the guide field, at the reconnection site.

By using purely hydrodynamic equations we have entirely neglected the energetic contribution from the magnetic field decrease across the slow magnetosonic shock.  With constant field strength the source of kinetic energy in our model must be decreasing field line length \citep[see][for an analysis of this]{Longcope2009}.  An initially bent field line relaxes to a straight line which is shorter by a factor $\cos(\Delta\theta/2)$, ignoring the current sheet thickness.    The magnetic energy decreases by this same fraction, provided the field strength does not change.  Had we accounted for the field decrease at the SS the fraction could be smaller still, so ours is a conservative estimate of the energy release.   On the other hand, the antiparallel limit ($\Delta\theta=180^{\circ}$) predicts a 100\% magnetic energy release from shortening alone, to which the field strength decrease can add nothing.

Figure \ref{fig:brat} shows that for  angles $\Delta\theta\ga120^{\circ}$ the post-shock field strength will decrease by more than one-third; this decrease is missing from our calculations.  The top panel shows that properly accounting for this decrease \citep[solid curve, for the full MHD solution of][]{Soward1982b} yields a post-shock temperature {\em lower} than that in a straight shock tube (dashed) which ignores it (albeit only slightly lower).  The reason that greater energy release results in lower temperature is that the flux tube is wider where it is weaker.  The extra magnetic energy released by weakening is more than fully consumed by work done expanding the tube.  The very similar post-shock temperatures in the two cases, in spite of the expansion in the first, leads us to believe that our shock-tube hydrodynamic model is adequate for studying thermal effects.  Neglecting magnetic energy release appears, paradoxically, to produce a slight {\em over}estimate of the temperature in the CF.

Our two-fluid modeling uses viscosity and thermal conductivity based on Coulomb collisions.  Notably absent is collisional resistivity which has played a central role in previous reconnection models.  Resistivity would have played its most most significant role generating the reconnection electric field within the diffusion region.  Our analysis has assumed this event to have occurred before $t=0$.  We have neglected direct energetic contributions from the resistive dissipation accompanying the electric field (i.e.\ Ohmic heating).  We neglect the contribution from this very small region in order to focus on the heating at the slow shocks.  

Nor would resistivity have significantly affected the shocks, had it been included.
Since the magnetic field bends only slightly at the slow shocks (see \fig\ \ref{fig:brat}), they carry little current and therefore account for little resistive dissipation.  In our shock tube treatment there is none at all.  More significant current will occur at the RDs, where the magnetic fields are bent following reconnection.  Including resistivity in the RD dynamics would diffuse these structures, replacing sharp bends with smoother corners.  In this case they are time-dependent structures rather than conventional steady shocks.  Diffusion would broaden the RDs, diminishing the direct Ohmic heating within, but would leave their global structure unaffected.  Thus we reach the conclusion that resistive dissipation makes a minor energetic contribution in skewed reconnection.

We find the use of collisional transport to be warranted for electron thermal conduction.  The rate of electron heating is low compared to the conductivity, so electron temperature gradients are always shallow.  The ion heating, directly from viscous dissipation, is far more rapid and leads to gradients at the limit of collisional (Chapman-Enskog) treatment.  This is related to the even more significant issue of the steep velocity gradients from the ion pressure and viscous stress.  These are also at the margin of fluid treatment; we found instances where the Chapman-Enskog treatment was unphysical, although only temporarily. Thus we expect non-fluid effects to be significant for ion dynamics within the sub-shock.  Previous studies have determined that such a treatment leads to a sub-shock thicker than that from fluid closures, by a factor in the range of two to four 
\citep{Greywall1975,Vidal1993}.  Aside from this thickened inner region, we expect our solutions to be approximately valid for cases of skew angle 
$\Delta\theta\la150^{\circ}$.

Some recent modeling of magnetic reconnection has focused on electric field from collisionless effects such as electron inertia or the Hall effect 
\citep{Biskamp1997,Shay1999,Birn2001}.  While most 
are two-dimensional studies of anti-parallel reconnection, some have included a guide field \citep{Pritchett2004,Hesse2004,Drake2006}.  The collisionless mechanisms generate an electric fields localized to a region of collisionless scale, such as $d_i=c/\omega_{pi}$.  The collisional scale of our solution is larger than this by
\be
  {\ell_{i1}\over d_i} ~\sim~ {v_{th,i}\over c}\,{\Lambda\over\ln\Lambda} ~~,
\ee
where $\Lambda=n\lambda_D^3\gg1$ is the plasma parameter and $\lambda_D$ is the Debeye length \citep{Nicholson1983}.  For the parameters we have used, 
$n_{e1}=10^{10}\,{\rm cm}^{-3}$ and $T_1=10^6$ K, the ion skin depth is 
$d_i=6\times10^{-4}\,\ell_{i1}$.  This means the mean-free path is effectively infinite compared to $d_i$, and the plasma appears collisionless on this scale.  Conversely, the collisionless effects occur on scales far smaller than all the structures in the outflow jets, including the slow magnetosonic shocks.

Collisionless effects would become significant were
flux transfer to occur within a region of collisionless scale.  This would enter our post-reconnection analysis as an initial gradient scale $\lambda\simeq d_i\ll\ell_{i1}$.  We have found that all initial conditions with $\lambda<\ell_{i1}$ converge to those we have analyzed within several ion collision times. This is within a distance $\Delta y\sim\ell_{i1}\beta_0^{-1/2}$ of the X-point itself, beyond which the solution would not differ from a purely collisional solution where $\lambda\sim\ell_{i1}$.  Thus we expect the large-scale appearance, including the slow shocks, to be similar even when collisionless processes are responsible for the flux transfer.

The initial relaxation of the velocity gradient in our fluid model is through viscous diffusion.  This would not be a valid approach were the initial gradient scale very small: 
$\lambda\ll\ell_{i1}$.  Ion-ion collisions would be ineffective over such small scales and the counter-directed parallel flows from the two RDs would interpenetrate, forming a central region of counter-streaming ion beams.  Such counter-streaming ions have been found in simulations \citep{Drake2009b} and spacecraft observations \citep{Gosling2005} when reconnection flux-transfer occurs on sub-collisional scales.  Calculations suggest that plasma instabilities will develop  to thermalize the ion distribution and form a hot, stationary, central region \citep{Parker1961}.  Since any such mechanism must satisfy the same conservation laws, the final structure will be identical to that found using two-fluid equations.  Only the thickness of the sub-shock will differ owing to the modification to the ion momentum diffusion (i.e.\ viscosity).

We believe our analysis is fairly robust and broadly applicable.  It models the dynamics following a sudden change to a field line's topology which was assumed to occur within a small region -- this is the flux transfer.  This topological change creates a mechanical disequilibrium of the field line causing it to shorten on the Alfv\'enic time scale.  The rapid shortening will drive compressive flows far faster than the sound speed (provided $\beta\ll1$).  It is the collision of these super-sonic compressive flows which generate the slow shocks and associated conduction fronts.  While we have studied the dynamics in a steady state, two-dimensional model, the same scenario has been identified in transient, three-dimensional reconnection \citep{Longcope2009,Guidoni2010}.  Even in the absence of a steady, coherent outflow jet, we expect Alfv\'enic retraction, slow shocks and conduction fronts to be inevitable features of fast magnetic reconnection.

\acknowledgements

DWL wishes to thank Terry Forbes for helpful discussions and to thank Jim Drake and the UMD Institute of for Electronics and Applied Physics for hosting his sabbatical leave.  The work was supported  by NSF and DOE under a joint grant.  SJB is supported by the NASA Living with a Star Program.


\end{document}